%% file: main.tex
\documentclass[fleqn,11pt]{wlscirep}

\usepackage[utf8]{inputenc}
\usepackage[T1]{fontenc}
\usepackage{import}
\usepackage{float}
\usepackage{caption}
\usepackage{subcaption}
\usepackage{tabularx}
\usepackage{color}
\usepackage{amsmath}
\definecolor{darkred}{rgb}{0.5,0.0,0.0}
\usepackage{scalerel}

\usepackage{geometry}

\title{The Coherence of US cities}

\author[1,2,*]{Simone Daniotti}
\author[3]{Matt{\'e} Hartog}
\author[1]{Frank Neffke}

\affil[1]{Complexity Science Hub Vienna, Vienna, 1080, Austria }
\affil[2]{Vienna University of Technology, Informatics, Vienna, 1040, Austria}
\affil[3]{Growth Lab, Harvard Kennedy School, Harvard University, Cambridge, MA 02138, USA}
\affil[*]{daniottisimone@gmail.com}

\begin{abstract}
Diversified economies are critical for cities to sustain their growth and development, but they are also costly because diversification often requires expanding a city's capability base. We analyze how cities manage this trade-off by measuring the coherence of the economic activities they support, defined as the technological distance between randomly sampled productive units in a city. We use this framework to study how the US urban system developed over almost two centuries, from 1850 to today. To do so, we rely on historical census data, covering over 600M individual records to describe the economic activities of cities between 1850 and 1940, and 8 million patent records as well as detailed occupational and industrial profiles of cities for more recent decades. Despite massive shifts in the economic geography of the U.S. over this 170-year period, average coherence in its urban system remains unchanged. Moreover, across different time periods, datasets and relatedness measures, coherence falls with city size at the exact same rate, pointing to  constraints to diversification that are governed by a city's size in universal ways. 

\end{abstract}
\begin{document}

\flushbottom
\maketitle
%
%
\thispagestyle{empty}

\section{Significance Statement}

\import{sections/}{significance}

\section{Introduction}

\import{sections/}{intro}
\section{Results}

\import{sections/}{results}

\section{Discussion}

\import{sections/}{discussion}

\section{Methods}

\import{sections/}{methods}

\bibliography{01_science_cities}  

\section*{Acknowledgements}
The authors thank the following persons for their helpful comments on earlier drafts of this paper:  Andrea Musso, Andr{\'e}s G{\'o}mez Li{\'e}vano, Johannes Wachs, S{\'a}ndor Juh{\'a}zs, Xiangnan Feng, James McNerney, Hillary Vipond. 

S.D. and F.N.  receive financial support from the Austrian Research Promotion Agency (FFG), project \#873927 (ESSENCSE). 

\section*{Author contributions statement}

All authors reviewed the manuscript.

\newpage
\appendix
\renewcommand{\thesection}{S\arabic{section}}
\counterwithin{figure}{section}
\counterwithin{table}{section}
\renewcommand{\theequation}{S\arabic{equation}} 
\setcounter{equation}{0} 
\renewcommand\thefigure{\thesection.\arabic{figure}}
\renewcommand\thetable{\thesection.\arabic{table}}

\section*{Supplementary information}

\import{sections/}{appendix}

\end{document}

%% file: sections/significance.tex
This study analyzes the nature and evolution of the economic coherence of cities. Much is known about the consequences of having a diversified urban economy, but we know little about how wide a range of activities a city can sustain. Here, we propose a measure of coherence that allows us to study changes in the breadth of economic activities in US cities over the course of 170 years. We find that, as the US economy transformed, economic activities became distributed across its urban system in ways that preserved coherence across cities. Moreover, coherence falls with city size at a rate that is constant across time periods and data sets. These findings suggest that the US urban system faced universal constraints along its development trajectory. This raises new types of questions about urban transformation and suggests that policymakers should take the constrained nature of urban transformation into consideration when devising interventions and plotting future development trajectories for their city.


%% file: sections/intro.tex
\begingroup
\renewcommand{\thefootnote}{}
\footnotetext{\textbf{Contribution Statement}

S.D. and F.N.: Conceptualization, Methodology, Writing – Original Draft.

S.D.: Investigation, Formal analysis, Visualization, Software.

F.N.: Supervision.

M.H. and S.D.: Data Curation.

All authors: Writing – Review and Editing.}
\endgroup

Diversification is pivotal to economic development and cities' capacity to generate prosperity for their inhabitants~\cite{younScalingUniversalityUrban2016,chongEconomicOutcomesPredicted2020,mazzarisiMaximalDiversityZipf2021}.
Diversified economies are less exposed to idiosyncratic sector-specific shocks \cite{rosenthalChapter49Evidence2004}, have a broader capability base from which to develop new economic activities \cite{hidalgoProductSpaceConditions2007,neffkeHowRegionsDiversify2011} and are better positioned to innovate \cite{ frenkenRelatedVarietyUnrelated2007,jacobsEconomyCities1969,feldmanInnovationCitiesSciencebased1999,mooreAnalysisJaneJacobs2017}. 
These and other consequences of diversification have been studied extensively. However, we know much less about how much diversity cities can manage, how this has changed over time, and how cities' capacity to do so is affected by their size. 
Here, we address these questions by leveraging large-scale micro-datasets that allow us to describe the evolution of the US urban system and the distribution of technological and economic activities across its cities almost from its inception to the present day. 
In particular, we study how \emph{coherent} the activity mix of a city is, in terms of the expected relatedness or technological proximity between two randomly sampled productive units in the city. The less coherent a city is, the broader the capability base required to support its activity mix will be. 
To do so, we develop a measure of coherence that is insensitive to economic classification systems and the exact measurement of relatedness. Studying the long-term evolution of the US urban system through this lens reveals that coherence falls with city size at a rate that stays the same across decades and datasets, suggesting the existence of universal constraints to diversification related to the size of a city.

A major challenge to studying the long-run transformation of urban economies is that economic activities and the classification systems to describe them change drastically over longer periods. Moreover, most existing approaches, which describe the breadth of activities in cities in terms of related and unrelated variety, tend to mechanically rise with city size~\cite{batheltRelatedVarietyRegional2022}, such that larger cities will, by construction, be more diversified. To address these challenges, we construct a measure of coherence that is \emph{a priori} unrelated to city size and insensitive to changes in classification systems. 

We use this coherence to study the long-run evolution of the US urban system between 1850 and today. To do so, we combine several large-scale micro-datasets, from historical census data that cover hundreds of millions of individuals in the 19th and 20th century to millions of patent records that describe the technologies used by US inventors between 1975 and 2020. These datasets cover different periods, different definitions of the US urban system -- which grew from 500 cities in 1850 to 900 cities today -- and different types of activities, such as occupations, industries, and technologies.   

Set against this heterogeneity, our analysis yields two surprising findings. First, the average coherence of cities in the US urban system remained constant in datasets that stretched over 170 years and across activity types. This stability is remarkable, given the drastic economic transformation that took place in this period, including the transition from agriculture to first manufacturing and then services, the rise and fall of the Rustbelt, and the emergence of today's technology hubs. Moreover, individual cities do undergo drastic structural change,  as in Detroit's rise and fall with the fortunes of its automotive industry and Boston's transformation from a port city to a city of higher education\cite{glaeser2005reinventing}. The fact that, in spite of such transformation, coherence remains unchanged therefore suggests that, as cities transform, they, on average, do so in a way that preserves the coherence among their (changing) activities along the way. Second, we uncover a universal relation between coherence and city size: the elasticity of coherence with city size is constant,  at about -4\%, across time periods and activity types. That is, moving from smaller to larger cities, the mix of activities broadens in a predictable way, such that doubling the size of a city translates roughly into a 4\% decrease in coherence. This holds not just true for the urban system as a whole, but also for the urban system on the West Coast of the US, which remained relatively disconnected from the eastern U.S. until the early 20th century and whose development we can trace in its entirety, starting from the mid 19th century. To help understand these patterns, we postulate that larger cities can maintain wider capability bases, which should allow them to develop less coherent activity portfolios, a logic we develop more formally in a model of collective learning in which cities' workers balance imitation and innovation.

Conceptually, our work relates to the framework of economic complexity\cite{hidalgoProductSpaceConditions2007,hidalgoBuildingBlocksEconomic2009a,balland2022new}. The literature on economic complexity assumes that economies mobilize capabilities to produce output. Although capabilities are treated as hidden, unobserved variables,   different products and services are generally assumed to require different capabilities.  This makes it costly to produce a wide variety of outputs because this will require a broad set of capabilities. Economies can save on the number of capabilities by focusing on sets of closely related activities. Accordingly, our coherence metric can be regarded as an attempt to assess the breadth of a city's capability base by analyzing how related different productive units in the city are to one another. The less related two randomly sampled units on average are, the more coherent the city and the broader its capability base will be.

The notion of coherence itself has been widely studied at the firm level in strategic management\cite{teeceUnderstandingCorporateCoherence1994,palich2000curvilinearity} as reviewed in\cite{robins2003measurement}, and more recently in research on economic complexity\cite{pugliese2019coherent,aufiero2024mapping}. Scholars in Evolutionary Economic Geography (EEG) then took the concept of coherence to the regional level. Our work relates most closely to papers in this latter tradition\cite{neffkeHowRegionsDiversify2011,essletzbichler2017relatedness}. However, we also take inspiration from information-theory-based metrics of diversity~\cite{raoDiversityDissimilarityCoefficients1982,vandamDiversityItsDecomposition2019,stirlingGeneralFrameworkAnalysing2007} to construct a readily interpretable metric that defines coherence as the expected relatedness -- for instance, in terms of cognitive or technological proximity -- between two randomly sampled productive units (e.g., workers) in a city. 

Finally, our study relates to a variety of academic fields that have documented striking relationships between diversification and economic outcomes in cities. First, economic geographers who study regional and urban diversity highlight its role in innovation~\cite{jacobsEconomyCities1969} by facilitating  Schumpeterian new combinations~\cite{schumpeterTheoryEconomicDevelopment1911}, agglomeration externalities~\cite{glaeserGrowthCities1992,marshallPrinciplesEconomics1890,porterCompetitiveAdvantageNations1990} and the path-dependent nature of regional diversification~\cite{boschmaEvolutionaryEconomicsEconomic1999,frenkenRelatedVarietyUnrelated2007,martinPathDependenceRegional2006,gWeaknessStrongTies1993}. 
Second, literature on urban scaling~\cite{bettencourtProfessionalDiversityProductivity2014,bettencourtOriginsScalingCities2013,bettencourtGrowthInnovationScaling2007,bettencourtDemographyEmergenceUniversal2020,younScalingUniversalityUrban2016} explores how economic activities scale with city size, concentrating in large cities~\cite{bettencourtUrbanScalingIts2010,schlapferScalingHumanInteractions2014,ballandComplexEconomicActivities2020a,durantonNurseryCitiesUrban2001}, and how occupational diversity relates to economic productivity and social network structures~\cite{bettencourtProfessionalDiversityProductivity2014,younSystematicStructurePredictability2016}.
However, these bodies of research tend to focus on the consequences of cities' activity mixes, not on what determines the breadth of the activity mix that cities can sustain in the first place. Moreover, most work analyzes cross-sections or short panels of cities in which activities are recorded in stable classification systems. As a consequence, they study cities and urban systems over years, instead of the decades and centuries over which their transformation typically unfolds.

%% file: sections/results.tex
\label{section:results}

\subsection{Data}

Our analysis draws from three different data sources. 
First, we use decennial US censuses for the period 1850-1940.
This dataset covers over 600M individual records and allows us to analyze changes in the occupational composition of between 550 and 900 cities in the U.S. in terms of 250 different occupations. 
Second, for the period between 2002 and 2022, we use data from the US Bureau of Labor Statistics (BLS) on employment by occupation for over 800 occupations in 350 metropolitan areas. 
For both data sources, we concentrate on occupations that are likely to produce tradable output whose geographic distribution is driven by  the availability of relevant capabilities, ignoring occupations that mainly cater to the demand of the local population, such as bakers, teachers, and nurses (see SI, sec.~\ref{sec:APP_tradable}). 
Third, we aggregate data from the US Patent and Trademark Office on over 8M patents between 1980 and 2020 to city-technology cells, distinguishing between 650 technological areas and 900 cities.  Together these datasets describe the occupational and/or technological composition of US cities between 1850 and today, except for the decades of 1890 for which a fire destroyed census records and of 1950, 1960 and 1970, for which neither comprehensive employment nor patenting information exists at the city level.  Details on cleaning and geocoding are provided in the \emph{Methods} section.

\subsection{Defining Coherence}

We define our metric of economic coherence in terms of the \emph{relatedness} between the economic activities in which productive entities in a city, such as workers, inventors or firms, are active. Relatedness has been measured in various ways and is often interpreted as a measure of cognitive or technological proximity. To show that our findings are not dependent on the exact definition of relatedness, we explore various relatedness metrics (see SI, sec.~\ref{sec:alternative}). First, using matched census records, we construct a measure of skill-relatedness~\cite{neffkeSkillRelatednessFirm2013} that connects occupations that exhibit exceptionally large labor flows between them. Second, in the BLS data, we derive measures of relatedness that express the extent to which two occupations are found in the same cities or industries. 
Third, in the patent data, relatedness expresses the degree to which two technologies co-occur on the same patents.

\begin{figure}[htbp]
    \centering
    \includegraphics[width=0.9\linewidth]{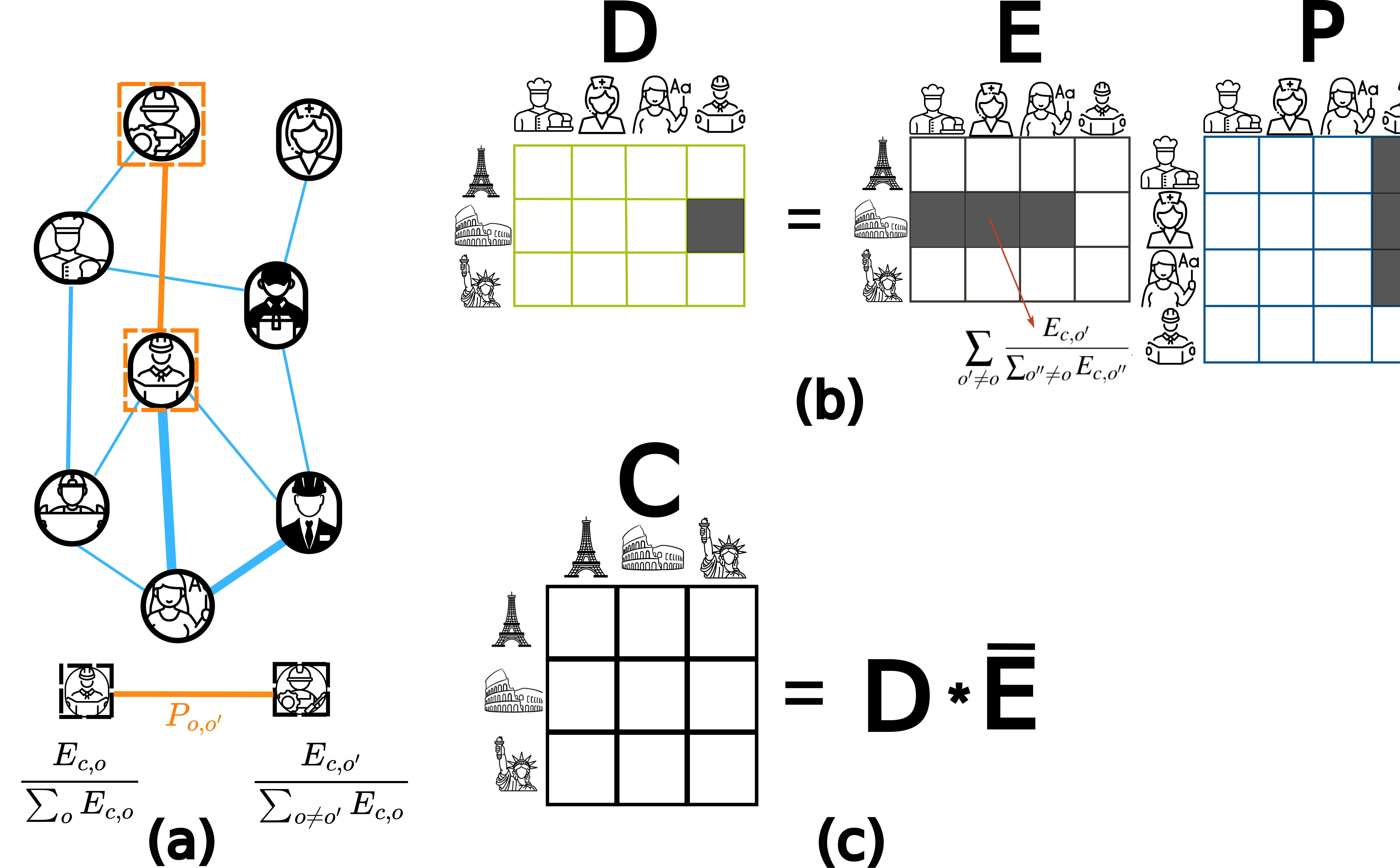}
    \caption{\textbf{Measuring coherence.} 
    \textbf{(a)} Proximity ($\bold{P}$). Stylized depiction of $(n_o \times n_o)$ proximity matrix as a network of $n_o$ activities (here: occupations) connected by edges that reflect their relatedness. 
    \textbf{(b)} Density ($\bold{D}$). The $(n_c \times n_o)$ density matrix reflects occupations' ``fit'' with each of the $n_c$ cities in our data. It is defined as the expected relatedness to all other occupations in the city, when the other occupations are sampled with probability $p=\frac{E_{c o}}{\sum_{l\neq o}E_{cl}}$, where $E_{co}$ captures the employment of city $c$ in occupation $o$. Dropping own-occupation contributions, element $D_{co}$ is calculated by multiplying a normalized row $c$ of the $(n_c \times n_o)$ employment matrix $\bold{E}$ with column $o$ of matrix $\bold{P}$, while omitting element $o$ from both vectors. 
    \textbf{(c)} Coherence. Estimated as the mean relatedness between workers in different occupations in the same city, coherence is calculated as $D\bar{E}$, where $\bar{.}$ indicates row-normalization by dividing all elements by corresponding row-sums.}
    \label{fig:matrices}
\end{figure}

Fig.~\ref{fig:matrices} illustrates how we estimate coherence. 
We first collect relatedness estimates for all pairs of activities in matrices $\bold{P}$. Next, we define coherence as the expected relatedness between two randomly sampled units, such as workers or patents, conditional on both units being sampled from the same city $c_1=c_2=c$. This involves two steps. 
The first step assesses how related an activity is to the rest of the urban economy. 
This measure is known as the activity's \emph{density} in the city\cite{hidalgoProductSpaceConditions2007,liEvaluatingPrincipleRelatedness2023}.
The second step averages this density across all activities.

To make matters concrete, we focus on the case of workers and their occupations. 
Elements $P_{o_1,o_2}$ now denote the proximity between occupations $o_1$ and $o_2$.  Furthermore, because the relatedness of an occupation to itself is undefined -- and to avoid coherence from picking up individual occupations' own geographical concentration  -- we condition this expectation on $o_1 \neq o_2$, where $o_1$ and $o_2$ denote the first and second sampled worker's occupations. 
This yields the following expression:
\begin{equation}\label{eq:coherence_1}
  C_{c_1,c_2} = \mathbb{E}(P_{o_1,o_2} | c_1=c,c_2=c',o_1 \neq o_2)  
\end{equation}

Elements of matrix $\bold{C}$ refer to pairs of cities. Coherence estimates are on $\bold{C}$'s diagonal, where the randomly sampled workers come from the same city. The off-diagonal elements contain the expected proximity between randomly sampled workers from different cities. These elements express how similar these cities are in terms of their activity mix. Expanding matrix $\bold{E}$ with a time dimension, such that it records the employment for a city in a given year in its rows, allows quantifying  urban \emph{transformation} as the expected proximity between randomly sampled workers from the same city, but at different points in time. The smaller this expected proximity, the more radically a city transforms.




Note that coherence is expressed in the same units as relatedness. Because relatedness definitions may differ across datasets and over time, we normalize coherence estimates by dividing them by a baseline that reflects the coherence of the US as a whole, i.e., the expected relatedness between workers randomly sampled from the entire US economy. The resulting ratio is independent of the unit in which relatedness is expressed and captures how much more or less related workers in the same city are than  workers in the US economy as whole.

\subsection{Structural Transformation of the US urban system}

The US population grew  from about 23 million inhabitants in 1850 to 132 million inhabitants in 1940 and 332 million inhabitants in 2022. With this growth in population, its economic structure underwent drastic transformation. Whereas in 1850, only about 15\% of the population was urban and about 60\% of the working population was employed in agriculture, by 1940, 56\% of the US population lived in cities and around 25\% of workers worked in manufacturing. Nowadays, 80\% of Americans live in urban areas and services have become the dominant sector, accounting for 70\% of all jobs.

\begin{figure}[t!]
    \begin{subfigure}[t]{0.32\textwidth}
        \centering
        \includegraphics[width=\textwidth]{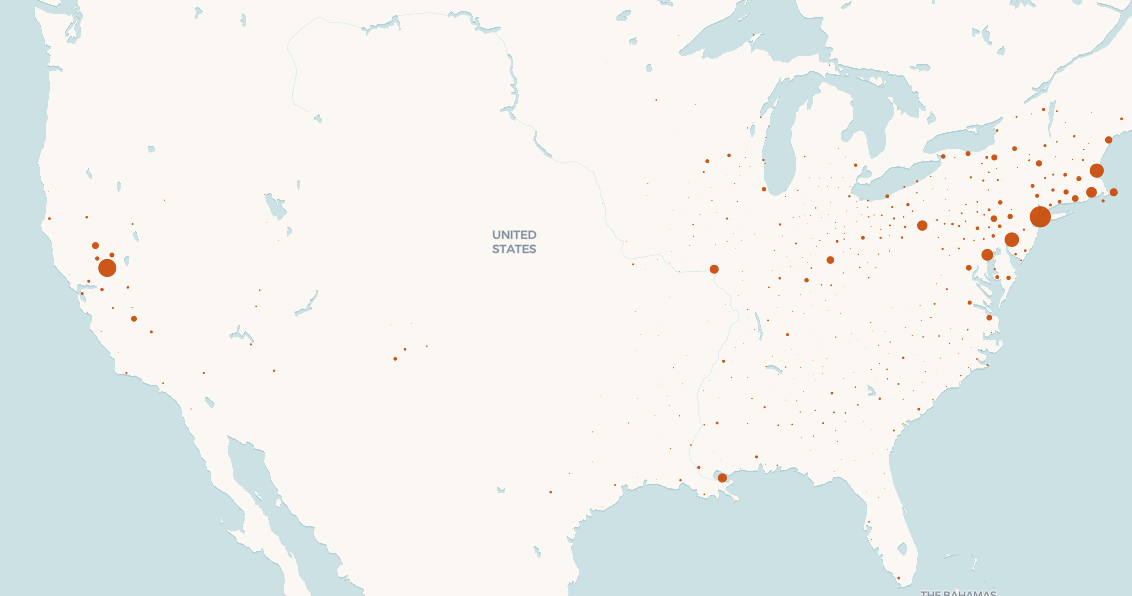}
        \caption{Employment in 1850}
        \label{fig:small1}
    \end{subfigure}
    \hfill
    \begin{subfigure}[t]{0.32\textwidth}
        \centering
        \includegraphics[width=\textwidth]{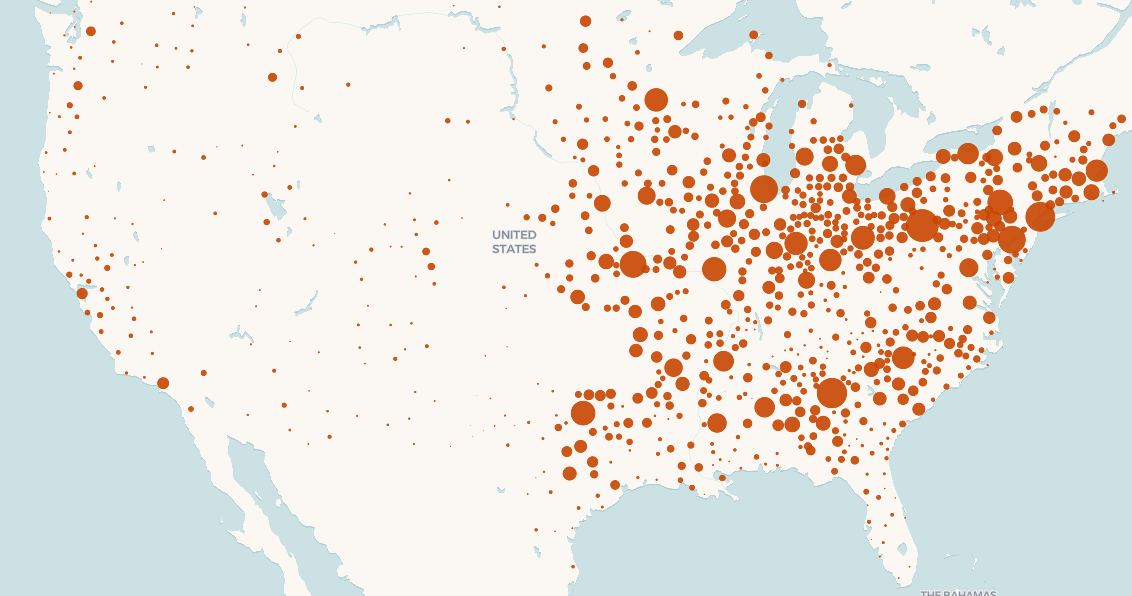}
        \caption{Employment in 1900}
        \label{fig:small2}
    \end{subfigure}
    \hfill
    \begin{subfigure}[t]{0.32\textwidth}
        \centering
        \includegraphics[width=\textwidth]{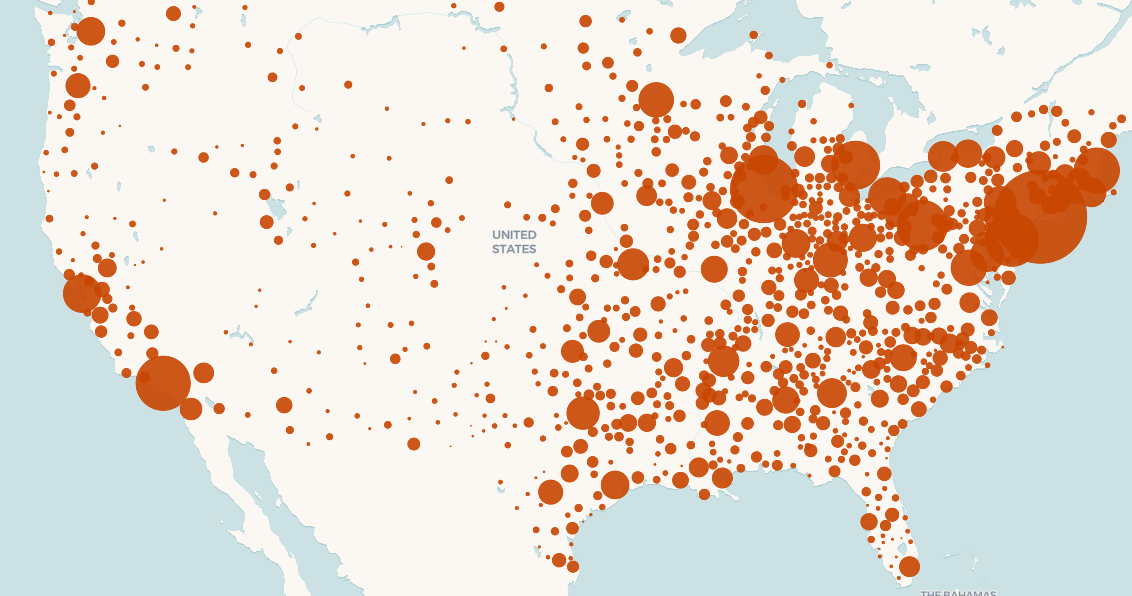}
        \caption{Employment in 1940}
        \label{fig:small3}
    \end{subfigure}
    \vspace{0.5cm} 
    \begin{subfigure}[t]{0.85\textwidth}
        \centering
        \includegraphics[width=\textwidth]{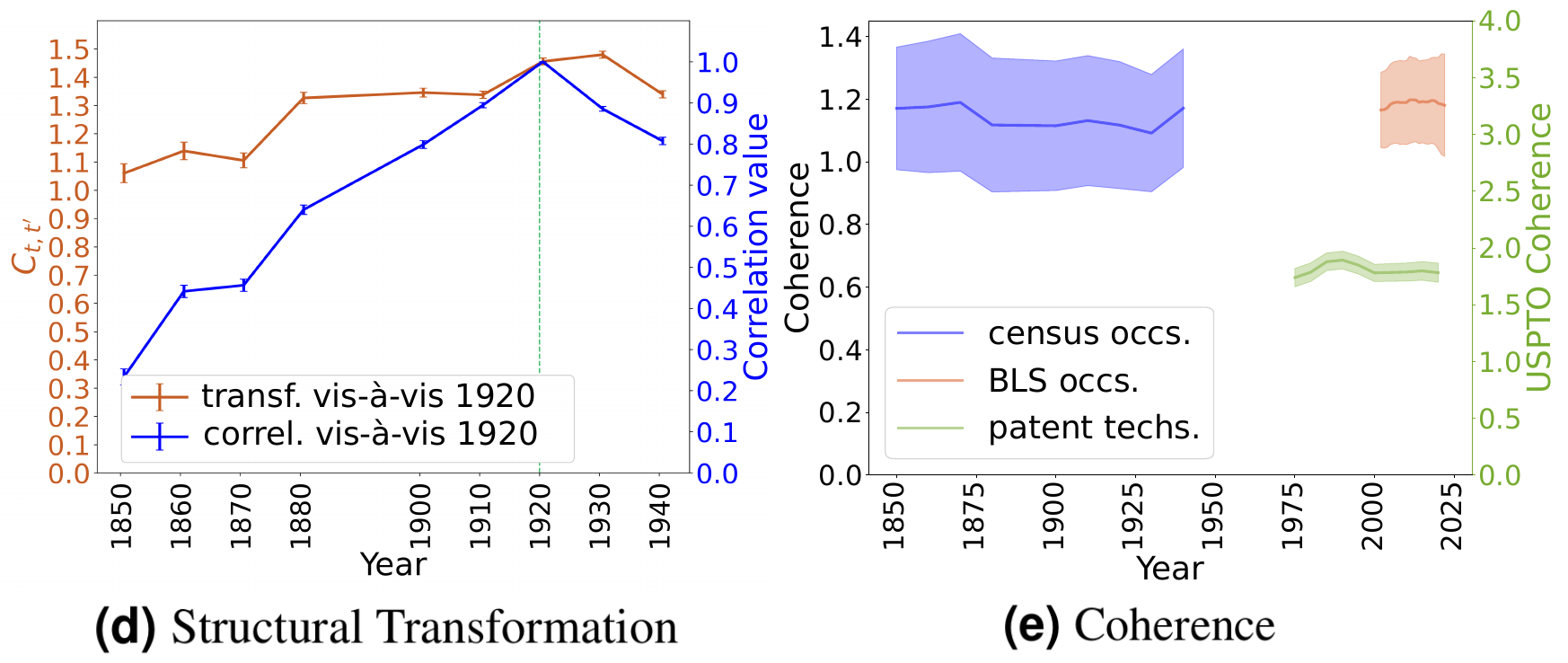}
    \end{subfigure}
    \vspace{-.4cm}
    \caption{\textbf{Structural Transformation at Constant Coherence.}} 
    Urban employment in the U.S. in \textbf{(a)} 1850, \textbf{(b)} 1900 and \textbf{(c)} 1940. Marker area is proportional to the number of employees in each city. The 1850-1940 period covers an important part of the formation and development of the US urban system in which the US went from a mostly rural to an advanced manufacturing and services economy.
    \textbf{(d)} Structural Transformation. Mean correlation between cities' occupational portfolios in different decades with their portfolio in 1920 in blue, and mean structural transformation vis-{\`a} vis 1920 in orange. Calculations are limited to cities  that exist in both decades and correlations  are calculated as 
    $\rho_{t,1920}^c=\text{corr}_{o\in O}\left(E_{oct},E_{oc1920}\right)$, where $E_{oct}$ denotes the employment in occupation  $o$, city $c$ and year $t$ with $O$ the set of occupations. Correlations are subsequently normalized by their country-level counterparts: $\rho_{t,1920}=\text{corr}_{o\in O}\left(E_{ot},E_{o1920}\right)$, where $E_{ot} =\sum_{c} E_{oct}$. Ratios $\frac{\rho_{t,1920}^c}{\rho_{t,1920}}$ are depicted along the vertical axis. Structural transformation is calculated as in eq.~(\ref{eq:structural}) and normalized by overall US-level transformation between $t$ and $1920$. 
    \textbf{(e)} Coherence. Mean coherence across US cities. Coherence is calculated as in eq.~(\ref{eq:coherence_1}) using occupation data from the census (1850-1940, blue line) and BLS (2002-2022, orange line), and patent data from the USPTO (1980-2020, green line). Values are normalized by dividing by system-level coherence. Confidence bands indicate 95\% confidence intervals.
    \label{fig:coherence}
\end{figure}

Despite such rapid transformation, Fig.~\ref{fig:coherence}e shows that cities' average coherence has remained constant for almost two centuries. The graph depicts the average coherence across cities in census data for the period 1850-1940 (blue line), BLS data for 2002-2022 (orange line), and patent data for 1980-2020 (green line). 
Across all datasets, coherence does not change in a statistically significant way.
Moreover, in terms of occupational coherence (blue and orange lines), there is no manifest change in average coherence between the towns and cities of 1850 and the urban areas and metropoles of today. This implies that, although cities substantially transformed their economies, they did so, maintaining a constant level of coherence. That is, as cities moved away from their past activities, they, on average, retained a constant level of ``compactness''.

\subsection{Coherence and city size}

\begin{figure}
    \centering
    
    \begin{subfigure}[t]{0.6\textwidth}
        \centering
        \includegraphics[width=\textwidth]{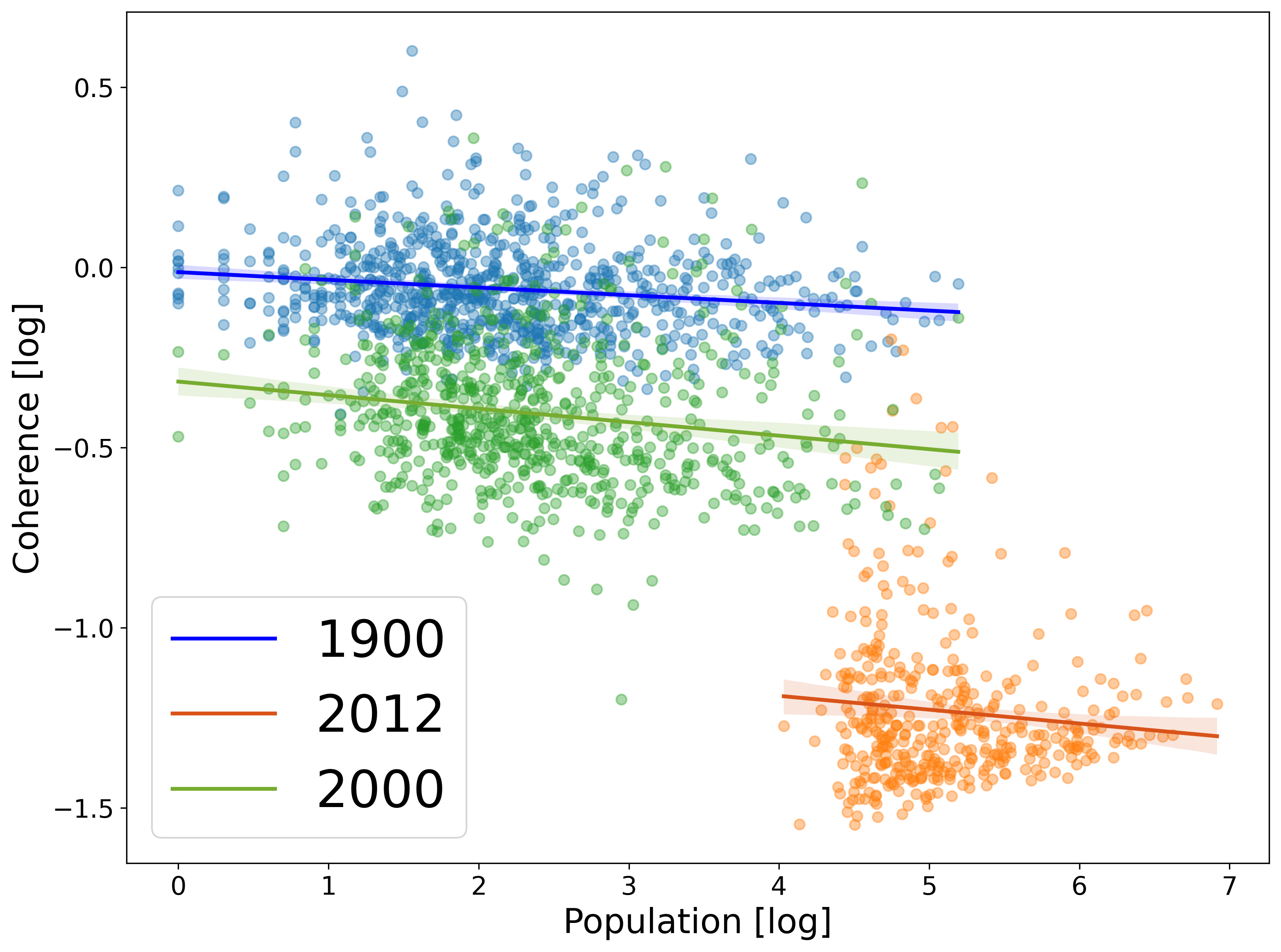}
        \caption{Coherence vs Population scatterplot}
        \label{fig:coh_vs_pop}
    \end{subfigure}
    \hfill
    \begin{subfigure}[t]{0.49\textwidth}
        \centering
        \includegraphics[width=\textwidth]{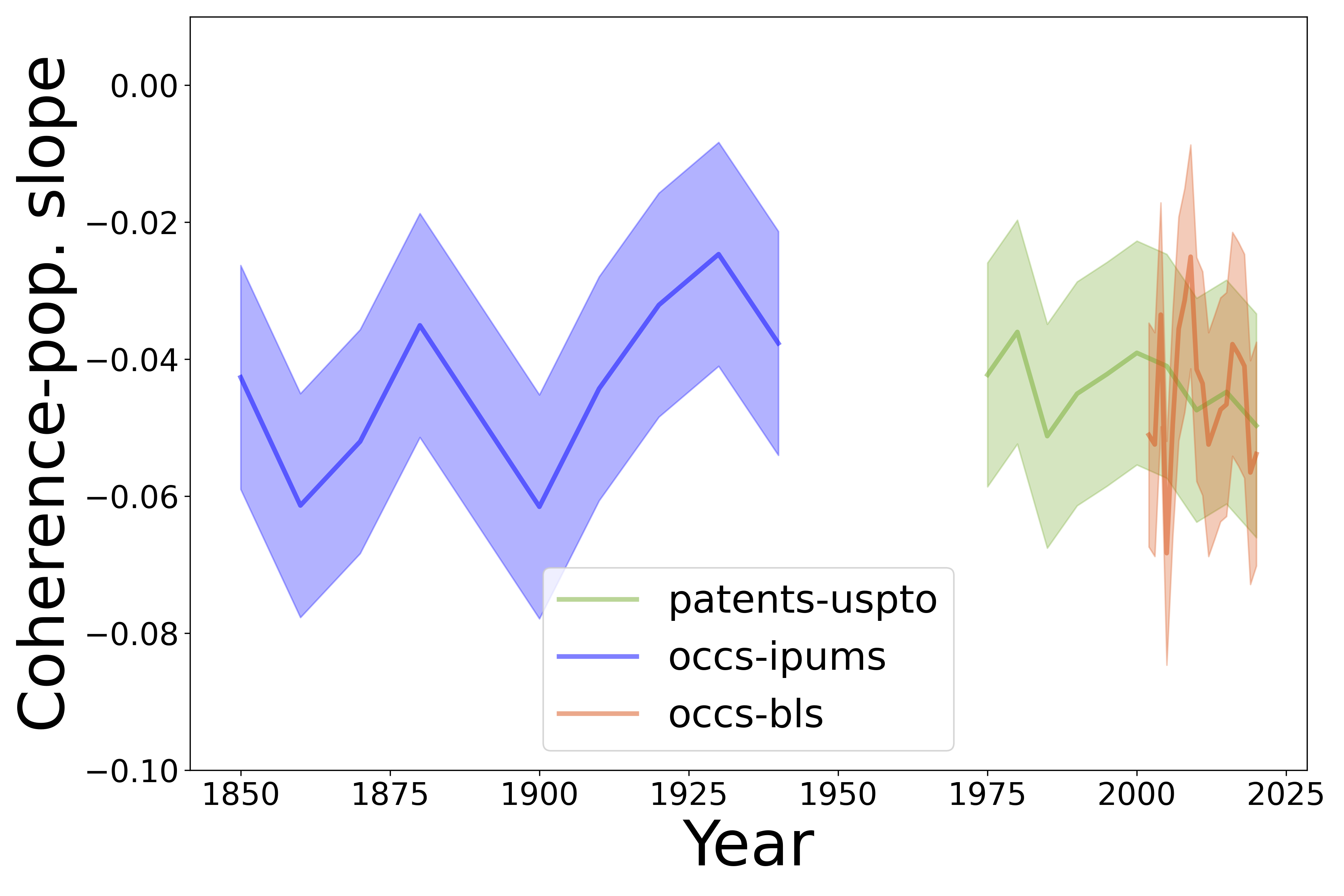}
        \caption{Coherence slope}
        \label{fig:slopes}
    \end{subfigure}
    \vspace{0.5cm} 
    \begin{subfigure}[t]{0.49\textwidth}
        \centering
        \includegraphics[width=\textwidth]{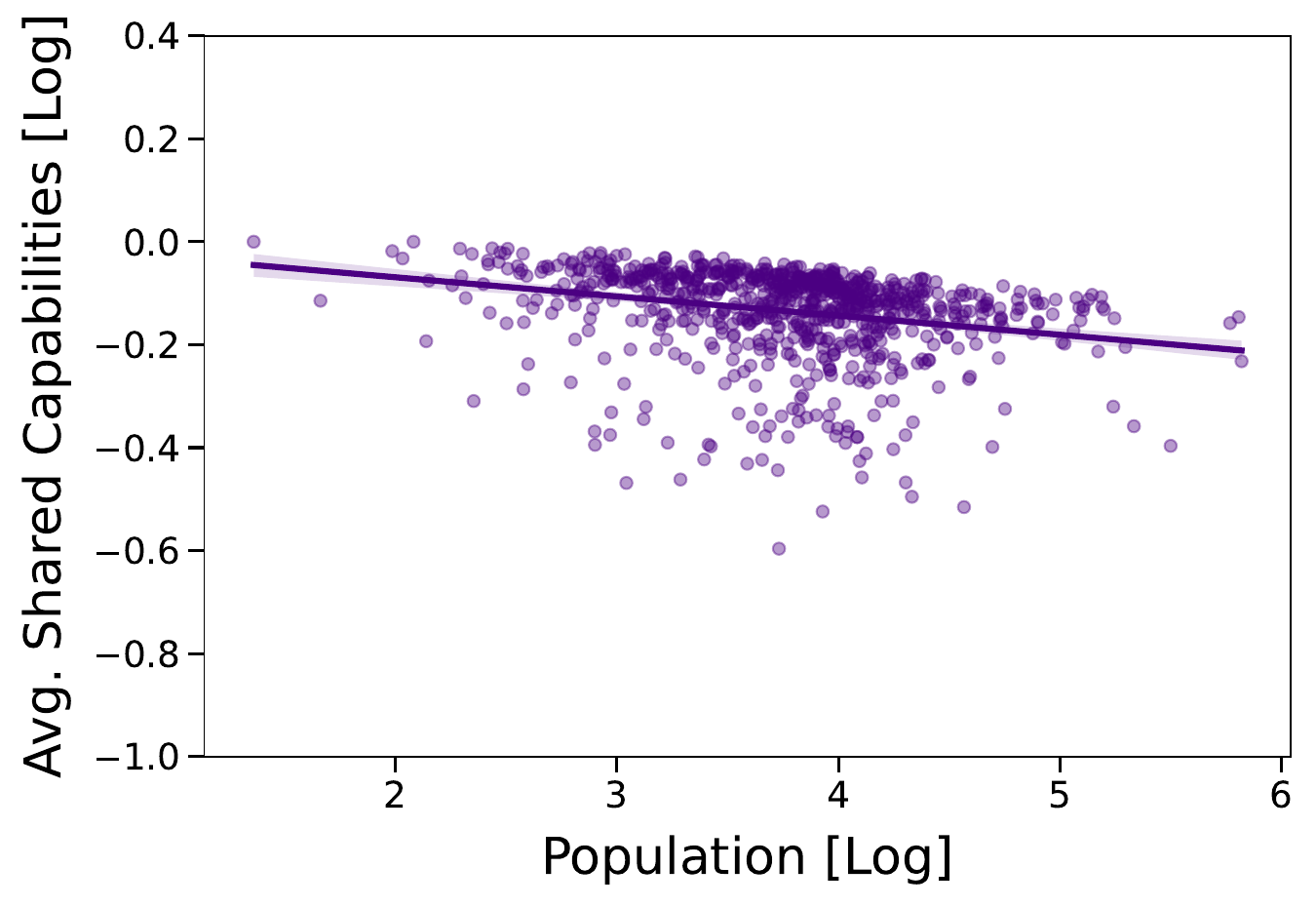}
        \caption{Shared Capabilities vs Population scatterplot}
        \label{fig:shared_skills}
    \end{subfigure}
    \caption{\textbf{Coherence versus Size}. \textbf{(a)} Scatter plot of coherence versus city size. Markers refer to cities in a specific dataset, blue: 1900 census, orange: 2012 BLS, green: 2000 patents. Lines represent best linear fits in each dataset. \textbf{(b)} Estimated elasticity of coherence with respect to city size. Colors are as in panel (a). Shaded areas reflect 95\% two-sided confidence intervals, based on robust standard errors. The null hypothesis of equal slopes cannot be rejected at any conventional level ($p=0.8$) in an equality-of-slopes Wald test.
    \textbf{(c)} Simulated coherence in a micro-simulation that balances innovation with imitation (SI, sec.~\ref{sec:sim}). City sizes are taken from the urban system of 1900 to mimic the blue scatter of panel a. The expected capability overlap, i.e., coherence, drops with an elasticity that depends on the parameter that governs workers' propensity to innovate (see Fig.~\ref{fig:sim}), which is  calibrated to the observed elasticity of -4\%, corresponding to an innovation propensity of 3\%. }
    \label{fig:coherence_vs_size}
\end{figure}

Although the urban system's average coherence remains constant, this does not necessarily hold true for individual cities. In general, diversity rises with city size~\cite{gomez-lievanoExplainingPrevalenceScaling2017}. Similarly, coherence falls with the city size. To study the relationship between coherence and city size, we regress the logarithm of coherence on the logarithm of the total number of workers in the city. Fig.~\ref{fig:coherence_vs_size}a shows that the relation between coherence and city size is downward sloping. More surprisingly,  the elasticity of coherence with respect to city size -- the slopes $\frac{\partial \log C_{cc}}{\partial \log E_c}$ in Fig.~\ref{fig:coherence_vs_size}a-\ref{fig:coherence_vs_size}b  -- are statistically indistinguishable across datasets and time periods (a Wald test for equality of slopes yields a p-statistic of 0.8) and close to -4\% (95\% confidence interval: [-4.4\%, -3.4\%]). That is, when city size doubles, coherence falls by approximately 4\%. 
As shown in Fig.~\ref{fig:logvar} of the SI, this contrasts with traditional measures of urban diversity, whose elasticities with respect to city size change markedly over the course of a century.

To help understand these findings, we develop a micro-simulation, where workers either imitate existing workers or innovate and develop new capabilities (see sec.~\ref{sec:sim} of the SI). In this context, a natural way to define coherence is the expected frequency with which two randomly drawn workers in a city will have the same capabilities. Fig.~\ref{fig:coherence_vs_size}c shows that by covering just two essential aspects of collective learning -- imitation and innovation -- this simple model is able to reproduce the functional form of the relation between coherence and city size. 

\subsection{West-Coast}
So far, our analysis has focused on the US urban system as a whole. While still relatively small, by 1850 the eastern part of this urban system had already become somewhat developed. The same is not true for the US West Coast. The population west of the Rocky Mountains amounted to just 300,000 people in 1850. In 1848, San Francisco had no more than 1,000 inhabitants, and the largest city in the region, Los Angeles, had 1,610 inhabitants, quickly surpassed in the following years by Sacramento.
Moreover, in most of the 19th century, the West Coast remained isolated from the rest of the US. Before the construction of the Panama Canal in 1914, ships had to round Cape Horn to travel between the Atlantic and Pacific Oceans and over land, mountain ranges acted as a significant barriers,  until the completion of the Transcontinental Railroad in 1869, which connected the East and West Coast and fueled rapid urban growth and economic development along its route. The US West Coast therefore offers  a unique opportunity to study how an urban system develops from scratch and then integrates into a larger, existing urban system.


\begin{figure}
    \begin{subfigure}[t]{0.32\textwidth}
        \centering
        \includegraphics[width=\textwidth]{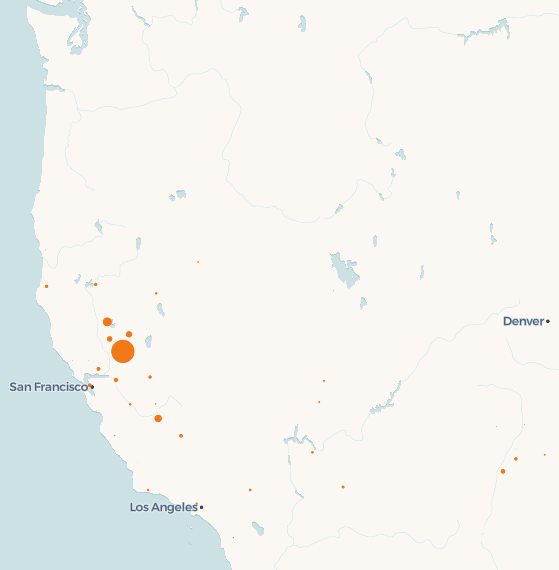}
        \caption{Employment West Coast 1850}
        \label{fig:small1_wc}
    \end{subfigure}
    \hfill
    \begin{subfigure}[t]{0.32\textwidth}
        \centering
        \includegraphics[width=\textwidth]{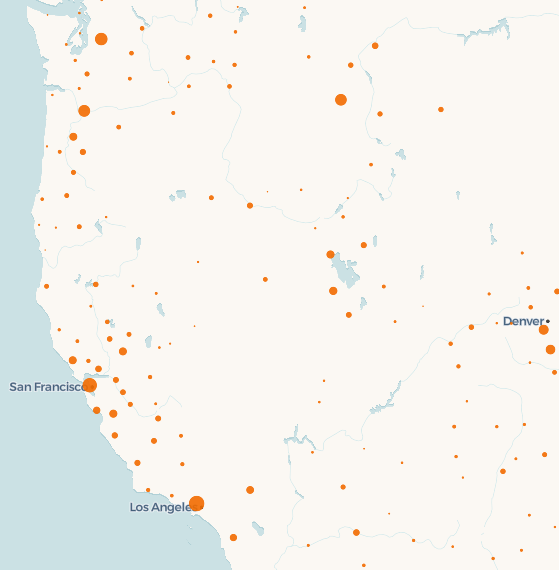}
        \caption{Employment West Coast 1900}
        \label{fig:small2_wc}
    \end{subfigure}
    \hfill
    \begin{subfigure}[t]{0.32\textwidth}
        \centering
        \includegraphics[width=\textwidth]{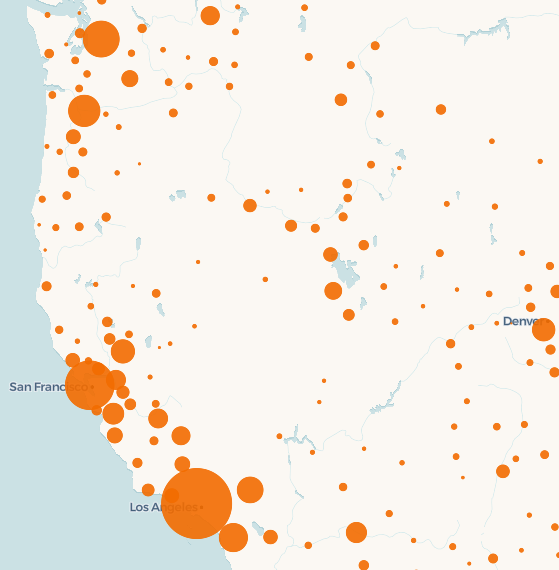}
        \caption{Employment West Coast 1940}
        \label{fig:small3_wc}
    \end{subfigure}
    \vspace{0.5cm} 
    \begin{subfigure}[t]{0.31\textwidth}
        \centering
        \includegraphics[width=\textwidth]{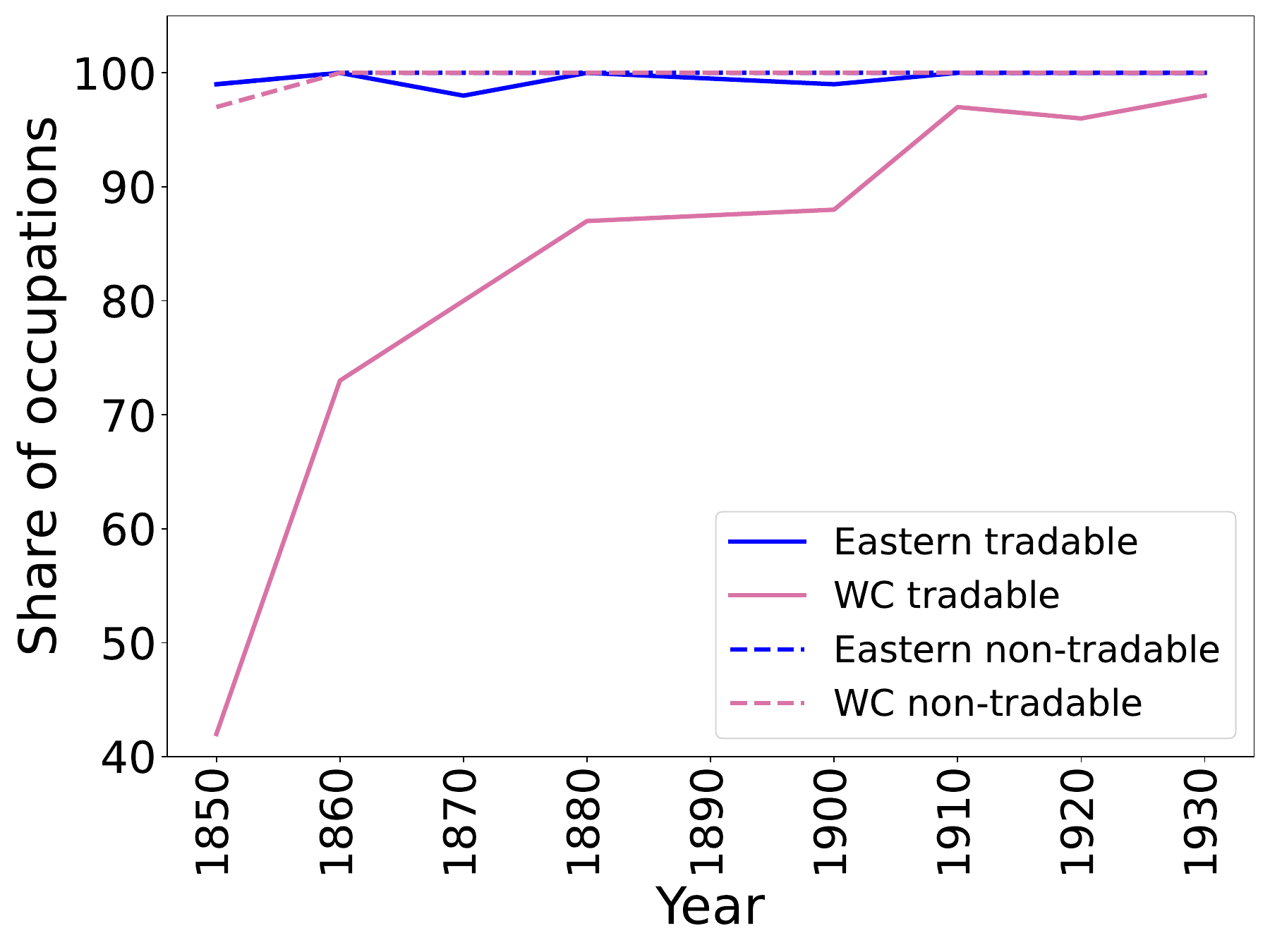}
        \caption{Occupational diversity}
        \label{fig:main}
    \end{subfigure}
    \hfill
    \begin{subfigure}[t]{0.32\textwidth}
        \centering
        \includegraphics[width=\textwidth]{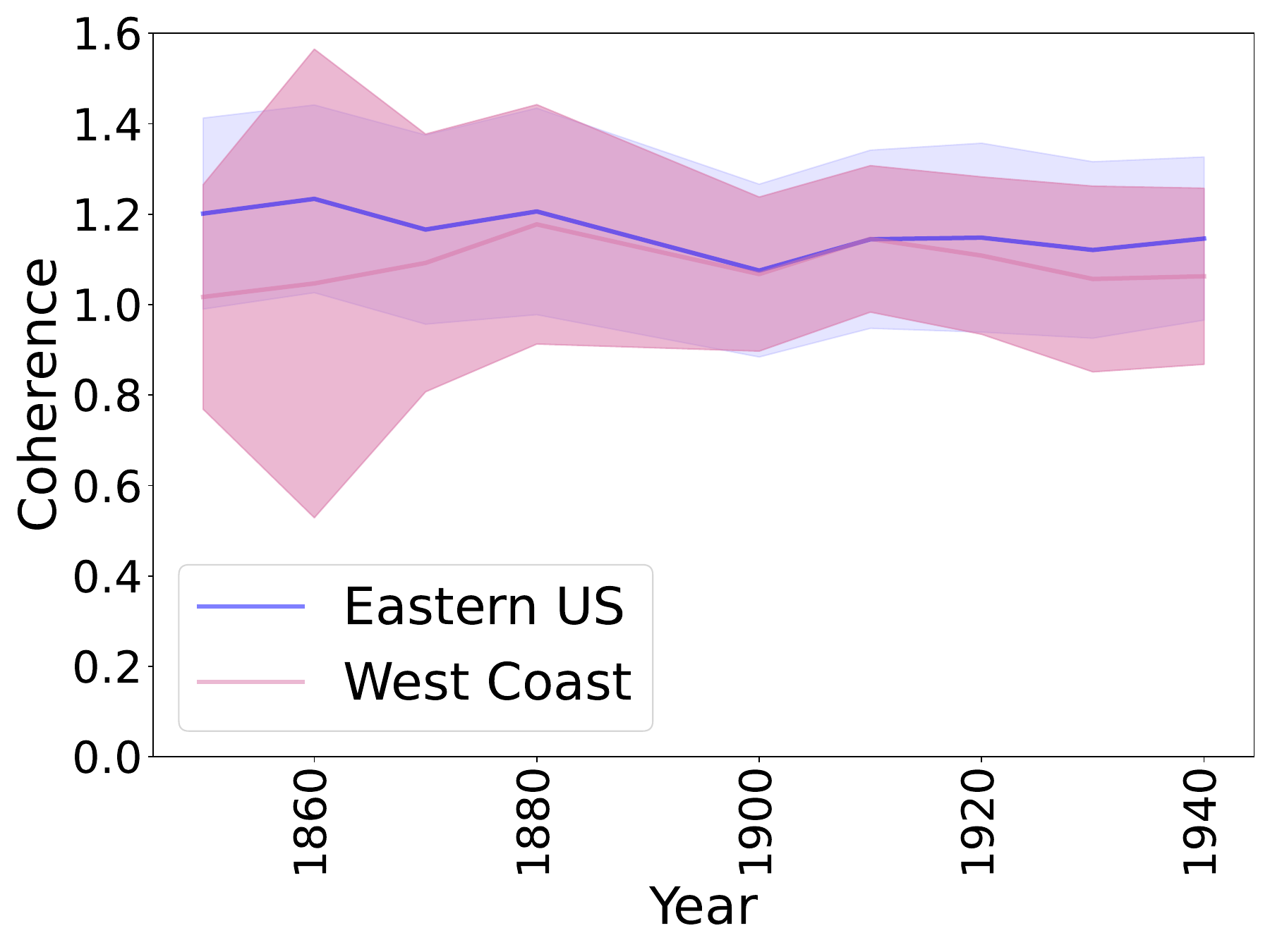} 
        \caption{Coherence}
        \label{fig:small2}
    \end{subfigure}
    \hfill
    \begin{subfigure}[t]{0.33\textwidth}
        \centering
        \includegraphics[width=\textwidth]{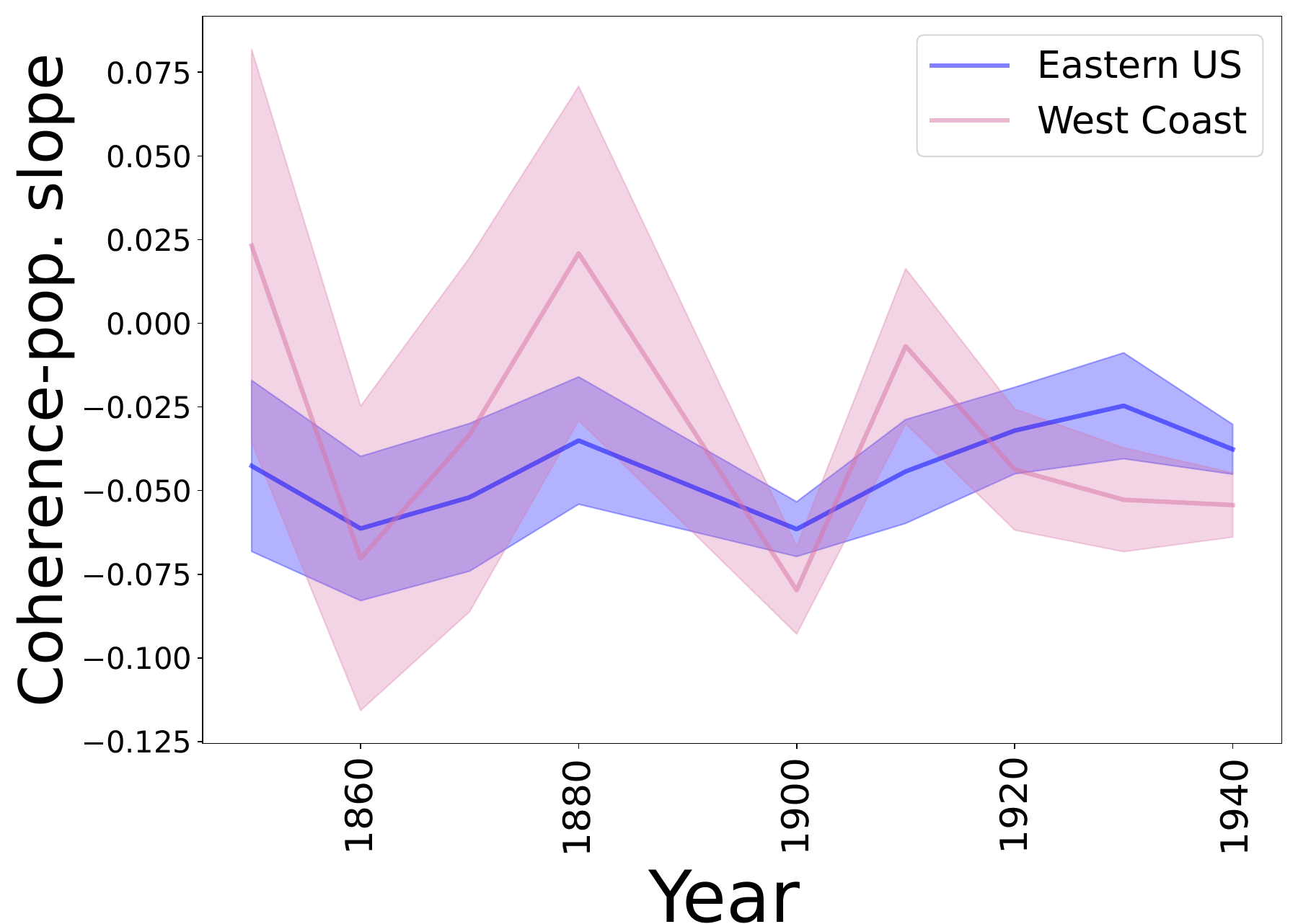} 
        \caption{Elasticity}
        \label{fig:westcoast_slopes}
        \end{subfigure}

    \caption{\textbf{US West Coast.} \textbf{(a-c)} Employment in West Coast cities in 1850, 1900 and 1940. In 1850, the largest city on the West Coast is Sacramento. After the gold rush, population growth shifts to 
    other cities, such as Portland, Seattle, San Francisco and Los Angeles.
    \textbf{(d)} Number of existing occupations as a share of all potential occupations in Eastern US (blue)  and on the West Coast (pink). Dashed lines refer to occupations in non-tradable activities, solid lines in tradable activities. Whereas occupations in non-tradables were already abundant on the West Coast in 1850,  in tradable activities,  less than half of all potential occupations had been developed by then. \textbf{(e)} Coherence. Mean coherence across US cities on the West Coast (pink) and in the Eastern US (blue). Coherence is calculated as in eq.~(\ref{eq:coherence_1}) and normalized by overall subsystem-level coherence. Confidence bands refer to 95\% confidence intervals. \textbf{(f)} Estimated elasticity of coherence with respect to city size. Shaded areas reflect 95\% two-sided confidence intervals, based on robust standard errors. By 1920, after some initial fluctuations,  West Coast cities exhibit the same level of coherence and elasticity of coherence with respect to city size as the remainder of the urban system.}  
    
\label{fig:westcoast_coherence}
    
\end{figure}

The structural transformation of the West Coast unfolded very fast. Fig.~\ref{fig:westcoast_coherence}d shows how quickly its cities  diversified: whereas, in 1850, they hosted just about 40\% of  existing occupations related to tradable activities, by 1900, this number had risen to close to 90\%. 
Even in this period of rapid diversification, average coherence of West Coast cities remained remarkably constant and, notably, at levels  indistinguishable from those in the eastern US (Fig.~\ref{fig:westcoast_coherence}e). Moreover, although the elasticity of coherence with respect to city size (Fig.~\ref{fig:westcoast_coherence}f) oscillates in the first 30–40 years -- possibly due to imprecise measurement in small populations  -- it thereafter converges rapidly to the same levels as observed in the rest of the U.S.


%% file: sections/discussion.tex

A diverse economy is of great importance for a city's capacity to innovate, grow and absorb adverse shocks. However, diverse economies require a range of different capabilities\cite{gomez-lievanoExplainingPrevalenceScaling2016}, which are often expensive to acquire and maintain. This begs the question of how broad a range of activities a city can sustain. To address this question, we have defined a city's coherence as the expected relatedness between randomly sampled productive units, e.g., workers or inventors, from the same city, while controlling for a nation-wide benchmark. We interpret this coherence as a proxy for the breadth of the city's capability base. This allowed us to address challenges inherent in long-term analyses of economic structures of cities, such as changing classification systems and distinguishing city-level change from broader economy-wide trends, while also focusing on fundamental changes as opposed to superficial shifts between closely related activities in a city's activity mix.

Applying this framework to data sets that describe the mix of economic activities in US cities over a 170-year time period uncovered important regularities. First, although the US urban system has undergone  substantial structural change, the coherence of cities within this system has, on average, remained remarkably stable. This suggests that cities' development trajectories are constrained: as cities transition from old activities to new ones, they, on average, maintain a constant level of internal coherence. 

Second, coherence decreases with city size at a universal rate. Across different time periods, relatedness measures and activity types (industries, occupations and patented technologies) the elasticity of coherence with respect to city size is constant at about -4\%. That is, coherence decreases by about 4\% with each doubling of a city's size, implying that  larger cities are able to support a broader set  of activities. The constancy of the point estimate of the relation between coherence and city size across periods and contexts suggests there may exist universal constraints that govern urban diversification. Interestingly, the estimated elasticity closely aligns with leading estimates of the urban wage premium in the U.S., according to which wages rise by around 5\% with each doubling of city size\cite{chauvin2017different}. Whether this is a coincidence or due to a connection between coherence and labor productivity is an interesting question for future research.

Third, after an initial turbulent period, cities on the West Coast settle into the same regularities as eastern US cities. The West Coast is an interesting case study, because our data describe its development more or less from the birth of its urban system, when geographical barriers initially still isolated it from the wider U.S.. In spite of this isolation and the rapid structural transformation it underwent, cities on the West Coast come to rapidly exhibit the same (constant) coherence and elasticity of coherence with respect to city size  as its counterparts east of the Rocky Mountains. This suggests that our findings may generalize to other urban systems. 



Our study has several limitations that can be tackled in future research. The first involves theoretical explanations for the functional form and  observed elasticity of -4\% for the relation between coherence and city size.
In the SI, Fig.~\ref{fig:sim}, we show that simple probabilistic models of imitation and innovation can reproduce our findings. This points to universality in collective learning as studied in the field of cultural accumulation\cite{kempeExperimentalDemonstrationEffect2014,klinePopulationSizePredicts2010}. However, other explanations may exist. One example is  regularities in the way division of labor deepens with city size and gives rise to new specializations\cite{bettencourtProfessionalDiversityProductivity2014}. Given the  invariance   across contexts and time of average coherence, as well as of its relation with city size, plausible candidates should be independent of technology and other aspects of societies that change on relatively short time horizons.  

A second limitation is the macro-level focus of our work on average proximity between a city's workers. However, what may matter most is whether workers can find a critical mass of closely related workers. Because large cities often consist of clusters of highly related activities\cite{porter2003economic}, workers may be able to  find a sizeable number of proximate workers even in cities with low coherence. This can be studied by looking at relatedness quantiles. For instance, one could calculate for each worker the 90th percentile of relatedness to other workers in the city. Cities with low levels of mean coherence but high levels of 90th percentile coherence would consist of disparate clusters of tightly related activities. This could explain why coherence falls with city size: although workers in large cities often find many workers in related activities, the wide variety of clusters they host lowers the average relatedness captured by our coherence metric. 

Finally, our findings have important implications for the broader discourse on regional development and growth policy~\cite{boschma2009evolutionary}. Such policy often focuses on fostering diversity in cities~\cite{frenkenRelatedVarietyUnrelated2007} and helping a city move into new economic activities to provide opportunities for growth and to avoid lock-in\cite{grabher1993weakness}. However, our analysis suggests that the breadth of activities a city can sustain is constrained by its size.  Similarly, the fact that, as cities transform, they maintain a constant level of coherence  suggests that there are structural constraints limiting the speed and trajectory of diversification. Therefore, diversification strategies should first benchmark a city's coherence against its size and analyze transformation trajectories that would allow the city to maintain its internal coherence.

%% file: sections/methods.tex
\subsection{Data}

\import{sections/}{data}

\subsection{Proximity}

The concept of \emph{proximity} between economic activities is central to research on Economic Complexity. In this field, economies are represented as networks of related activities\cite{hidalgoProductSpaceConditions2007,neffkeHowRegionsDiversify2011,hidalgoPrincipleRelatedness2018}, where relatedness captures the degree to which different activities require similar capabilities. When cities develop related activities, they can support a wide variety of such activities with a limited set of capabilities. In this context, our coherence measure can be viewed as a way to quantify a city's (lack of) diversity, not in terms of its activities, but of its capabilities. This approach resonates with research on diversity in ecosystems, which often distinguishes between variety, balance and disparity\cite{vandamDiversityItsDecomposition2019}. We discuss the relation between coherence and the metrics in this literature in the SI, section~\ref{sec:Hillnumbers}.

Relatedness can be measured in various ways \cite{liEvaluatingPrincipleRelatedness2023}. For the census data, we base relatedness on labor flows, i.e., on a count of how many individuals move from one occupation to another between two consecutive census waves. To be precise, we assess to what extent the labor flows between occupation $o$ and $o'$ are surprisingly large, using Pointwise Mutual Information as a metric of surprise:

\begin{equation}
    \operatorname{PMI}\left(p_{o o'}\right)=\log \left(\frac{p_{o o'}}{p_o p_{o'}}\right),
\end{equation}
where $p_{o o'}$ is the joint probability that an individual moves from occupation $o$ to occupation $o'$ and  $p_{o}$ and $p_{o'}$ are the marginal probabilities of moving out of occupation $o$ and into occupation $o'$. We estimate these probabilities by the obsersved relative frequencies. For instance, we  estimate $p_{o o'}$ as $\hat{p}_{o o'} = \frac{F_{o o'}}{\sum_{k,l}F_{k l}},$ where $F_{o o'}$ is the observed labor flow from occupation $o$ to $o'$.

Often, authors draw a sharp distinction between relatedness and unrelatedness\cite{muneepeerakul2013urban,liEvaluatingPrincipleRelatedness2023}. Following recommendations in this literature, we define proximity as:
\begin{equation}\label{eq:est_prox}
    P_{oo'} =
    \begin{cases}
        \widehat{PMI}(\frac{F_{oo'}}{\sum_{k,l}F_{k l}}) & \text{if } \widehat{PMI}(\frac{F_{oo'}} {\sum_{k,l}F_{k l}})>0, \\
        0 & \text{otherwise}
    \end{cases}
\end{equation}
where $F_{oo'}$ is the mean of the labor flow between two occupations $o$ and $o'$ across all pairs of consecutive census waves and $\widehat{PMI}$ is estimated using the Bayesian approach in\cite{vandamInformationtheoreticApproachAnalysis2023}. This sets all negative elements of matrix $\bold{P}$ to zero. Furthermore, because the relatedness of an activity to itself is ill-defined, diagonal elements of proximity matrices are ignored in the definition of coherence (see below). 
 
Replacing flows by co-occurrences, we  can also derive estimates of proximity from the frequency with which two occupations co-occur in the same industry or city. We use city-level co-occurrences as an alternative proximity metric in our census data and  city and industry co-occurrences to produce two different proximity metrics in the BLS data. In patent data, we calculate proximity from the frequency with which technology codes co-occur on the same patent. 
Our results prove remarkably robust to changes in the definition of proximity (sec.~\ref{sec:alternative} of the SI).

\subsection{Defining Coherence}

We define coherence as the expected proximity between two randomly sampled workers, conditional on the workers being from the same city and employed in different occupations. Breaking down the calculation of coherence into two  steps helps connect the coherence metric to the literature on economic complexity\cite{hidalgoProductSpaceConditions2007,hidalgoEconomicComplexityTheory2021,balland2022new}. This is illustrated in Fig.~\ref{fig:coherence}. First, we calculate the weighted average proximity of a given occupation, $o$ to all other occupations in the city $c$. This quantity is closely related to what the economic complexity literature\cite{liEvaluatingPrincipleRelatedness2023} refers to as $o$'s \emph{density}, $D_{oc}$, in city $c$:

\begin{equation}\label{eq:density}
D_{oc} =\sum_{o' \neq o} Pr(o_2=o'|c_1=c,c_2=c,o'\neq o) P_{oo'},
\end{equation}
which can be estimated as:
\begin{equation}\label{eq:density_estimate}
\hat{D}_{oc} =\sum_{o' \neq o} \frac{E_{o'c}}{\sum_{o'' \neq o}  E_{o''c}} P_{oo'}
\end{equation}

Coherence, a city-level variable, is now simply the expected density across all occupations. To calculate this, we construct the following matrix:
\begin{equation}\label{eq:coherence}
\begin{aligned}
  C_{c_1,c_2} &= \mathbb{E}(P_{o,o'} | c_1=c,c_2=c',o_1 \neq o_2)  \\
  &= \sum_{o} Pr(o_1=o|c_1=c,c_2=c') \sum_{o'} Pr(o_2=o'|c_1=c,c_2=c',o\neq o')  P_{oo'},
\end{aligned}
\end{equation}
which can be estimated as the employment-weighted average density:
\begin{equation}\label{eq:coherence_estimates}
\begin{aligned}  
  \hat{C}_{c_1,c_2} &= \sum_o \frac{E_{oc}}{\sum_{o''}  E_{o''c}}  \sum_{o' \neq o} \frac{E_{o'c'}}{\sum_{o'' \neq o}  E_{o''c'}} P_{o,o'} \\
  &= \sum_o \frac{E_{oc}}{\sum_{o''}  E_{o''c}}  \hat{D}_{oc'},
\end{aligned}
\end{equation}
where $E_{co}$ denotes the number of workers employed in occupation $o$ and city $c$.
A city's coherence is found on the diagonal of matrix $\bold{\hat{C}}$, which contains estimates of the expected proximity between workers that were sampled from the same city. The average coherence across the urban system, plotted in Fig.~\ref{fig:coherence}e, is calculated as the weighted average of the diagonal elements of $\bold{\hat{C}}$, using cities' overall employment as weights. Next, we rescale this estimate by dividing by the analogous quantity for the US economy as a whole. That is we combine all cities other than $c$ into one unit such that the calculations in eq.~(\ref{eq:coherence}) yield a scalar. Confidence intervals are based on estimates of the standard errors of eq.~(\ref{eq:coherence_estimates}), which are calculated as follows:

\begin{equation}
    \sigma(\hat{C}_{c_1,c_2}) = \sqrt{
        \sum_{o} \sum_{o' \neq o} 
        \left[
            \frac{E_{oc_1}}{\sum_{o''} E_{o''c_1}}  
            \frac{E_{o'c_2}}{\sum_{o'' \neq o} E_{o''c_2}}
        \right]^2 
         \sigma(p_{oo'})^2
        },
\end{equation}
where $\sigma(p_{oo'})$ is the Bayesian estimate of the standard deviation of the proximity between occupations $o$ and $o'$\cite{vandamInformationtheoreticApproachAnalysis2023}.

The off-diagonal elements of $\bold{\hat{C}}$ also have a useful interpretation: they can be regarded as estimates of the proximity between two cities. We use this to quantify the amount of \emph{structural transformation} a city undergoes. To do so, we include into matrix $\bold{\hat{C}}$ observations for the same city at different points in time. This yields elements that estimate the expected proximity between two workers that were sampled from the same city, but in different years:    
\begin{equation}
\label{eq:structural}
  \hat{C}_{c_1,c_2}^{(t,t+\tau)} = \sum_o \bar{E}_{oc_1}^{(t)} \sum_{o' \neq o} \frac{E_{o'c_2}^{(t+\tau)}}{ \sum_{o'' \neq o}  E_{o''c_2} ^{(t+\tau)}} P_{oo'},
\end{equation}
where $\bar{.}$ indicates row-normalization by dividing by row-sums. Values $\hat{C}_{c_1,c_2}^{(t,t+\tau)}$, normalized by the estimated average coherence in 1920, are shown in orange in Fig.~\ref{fig:coherence}d.


%% file: sections/data.tex
Our analysis is based on three different datasets: US census records between 1850 and 1940, Bureau of Labor Statistics (BLS) data between 2002 and 2022 and United States Patent and Trademark Office (USPTO) data between 1980 and 2020.


\paragraph{US census records.}

Census data are provided by the Integrated Public Use Microdata Series known as IPUMS \cite{rugglesIPUMSUSAVersion2021}. This dataset has approximately 650 million records of responses to Census inquiries for every resident in the United States during the years 1850, 1860, 1870, 1880, 1900, 1910, 1920, 1930, and 1940 (the 1890 records were destroyed in a fire). These records include essential details for our analysis such as each individual's name, occupation, industry, year and state of birth and place of residence. We focus on the working population, which we define as individuals aged 15 to 65 for whom an occupation of employment is recorded. We exclude individuals in occupational categories ``Unknown'' or related to agriculture, given that the latter are not part of the urban economy.

Individuals were geocoded and linked across census waves  by \cite{protzer2024new}. We aggregate these data to the level of cities using point-in-polygon merges, where polygons are the metropolitan and micropolitan areas defined by the US Census bureau in its TIGER/Line Shapefiles (\url{https://www.data.gov/}). 
 We use the urban shapes for the year 2020, projecting them backward in time to maintain the same spatial definitions of cities.
The result is a dataset with employment for approximately 250 occupations in between 550 US cities in 1850 and 900 cities in 1940.

\paragraph{BLS data.}
The BLS data are taken from the BLS Occupational Employment Statistics (OES) tables, available at \url{https://www.bls.gov/oes/tables.htm}. These tables record the number of employees in approximately 800 occupations across about 350 US metropolitan areas. Furthermore, we use the BLS' industry-occupation matrix, which records the number of employees in occupation-industry cells to calculate relatedness between occupations from their co-occurrence across  industries.

\paragraph{Patent data.} The USPTO dataset are obtained from PatentsView, \url{https://www.uspto.gov/ip-policy/economic-research/patentsview}. We focus on patents granted by the USPTO between 1980 and 2020, geocoding US inventors based on their places of residence through point-in-polygon merges to TIGER/Line Shapefiles. 
We aggregate patents to the city-technology level, distributing each patent proportionally to the share of its inventors in each cell. This yields a dataset of (fractional) patent counts for approximately 650 technologies in 900 cities.

\paragraph{Tradable and nontradable occupations.} An important distinction exists between economic activities that cater to the need of a city's own population and those whose output is traded with other cities. The former are called ``nontradable'' and include occupations such as bakers, school teachers, doctors, retail workers, etc.. Demand for nontradable activities is driven mostly by the size of the local population and its purchasing power, such that  employment shares in nontradable 
 occupations are very similar across cities. In contrast, tradable activities depend on the city's productivity in these activities. Examples include  manufacturing activities, but also  services sold to inhabitants of other cities, such as investment banking, research and development or higher education.  

Unlike a city's nontradable activities, which tradable activities a city can develop depends on its capability base. Consequently, the capability base of a city is best reflected in its tradable activities and we therefore drop nontradable activities from our analysis of a city's coherence. In patent data, we consider all activities  tradable, given that patents protect inventions on the entire US market. When analyzing occupations, we leverage the fact that nontradable activities essentially follow population and calculate for each occupation how closely its distribution across cities follows the distribution of the US population. That is, we calculate the correlation between two vectors, $\vec{e_{o}}$, whose elements, $E_{oc}$, contain the employment of occupation $o$ in city $c$ and $\vec{e}$, whose elements, $E_{c.}=\sum_o E_{oc}$, describe city $c$'s overall employment as a proxy for its population. This yields the following nontradability score: ${NT}_o = corr\left(\vec{e_{o}},\vec{e} \right)$.

Fig.~\ref{fig:tradables_si} in the SI shows that coherence estimates rise the more we limit the analysis to tradable occupations in census and BLS data. In the census data, we observe a sharp transition after removing $70\%$ of the least tradable occupations.This shows that coherence is mostly driven by tradable occupations. Although in the BLS data, we do not find a specific transition point, we observe the same strong relation between tradability and coherence.
We therefore define tradable occupations in both census and BLS data as occupations with ${NT}<0.7$.

%% file: sections/appendix.tex
This document provides  supplementary information to the paper ``The Coherence of US Cities.'' Sec.~\ref{sec:APP_tradable} classifies jobs into tradable and nontradable occupations using a nontradability index. Sec.~\ref{sec:alternative} explores  different methods to calculate occupational relatedness and replicates the paper's main results using the resulting alternative measures. Sec.~\ref{sec:Hillnumbers} discusses the relation between our coherence measure and the notions of variety, balance and  disparity as developed in research on diversity in ecosystems. Sec.~\ref{sec:sim} explores the relation between city size and coherence in a micro-simulation where productive units choose between innovating new activities or imitating existing activities.

\section{Tradable versus nontradable occupations}\label{sec:APP_tradable}

We have focused our analysis on occupations in tradable activities (``tradable occupations''). The rationale is that, if coherence is to assess  the breadth of a city's capability base, demand for the occupations we use to measure coherence should be driven by a city's capabilities. This is not necessarily the case for occupations that cater to local demand, whose size mainly reflects the size and preferences of a city's population, not its capabilities. The dependence on local customers also constrains the growth of these occupations to the local demand. In contrast, occupations that are used in activities that can be exported to other cities (``tradable occupations'') can grow independently of the size of the local market. Their growth will instead be determined by the city's productivity in the activity, in other words, by the nature of the city's capability base. Tradable occupations therefore capture what a city is good at, making them more informative of a city's coherence than nontradable occupations. 

To distinguish nontradable from tradable occupations we leverage the fact that nontradable occupations follow local demand. Consequently, their spatial distribution across cities will closely mimic the spatial distribution of population. Following this logic, in the main text, we defined a nontradability index as  ${NT}_o = corr\left(\vec{e_{o}},\vec{e} \right)$, where $\vec{e_o}$ is a vector that describes the distribution of workers in occupation $o$ across cities and $\vec{e}$ a vector describing the distribution of all workers across cities. The higher this correlation, the more the occupation's spatial footprint mimics the spatial footprint of the US population.

\begin{figure}
    \centering
    \includegraphics[width=0.4\linewidth]{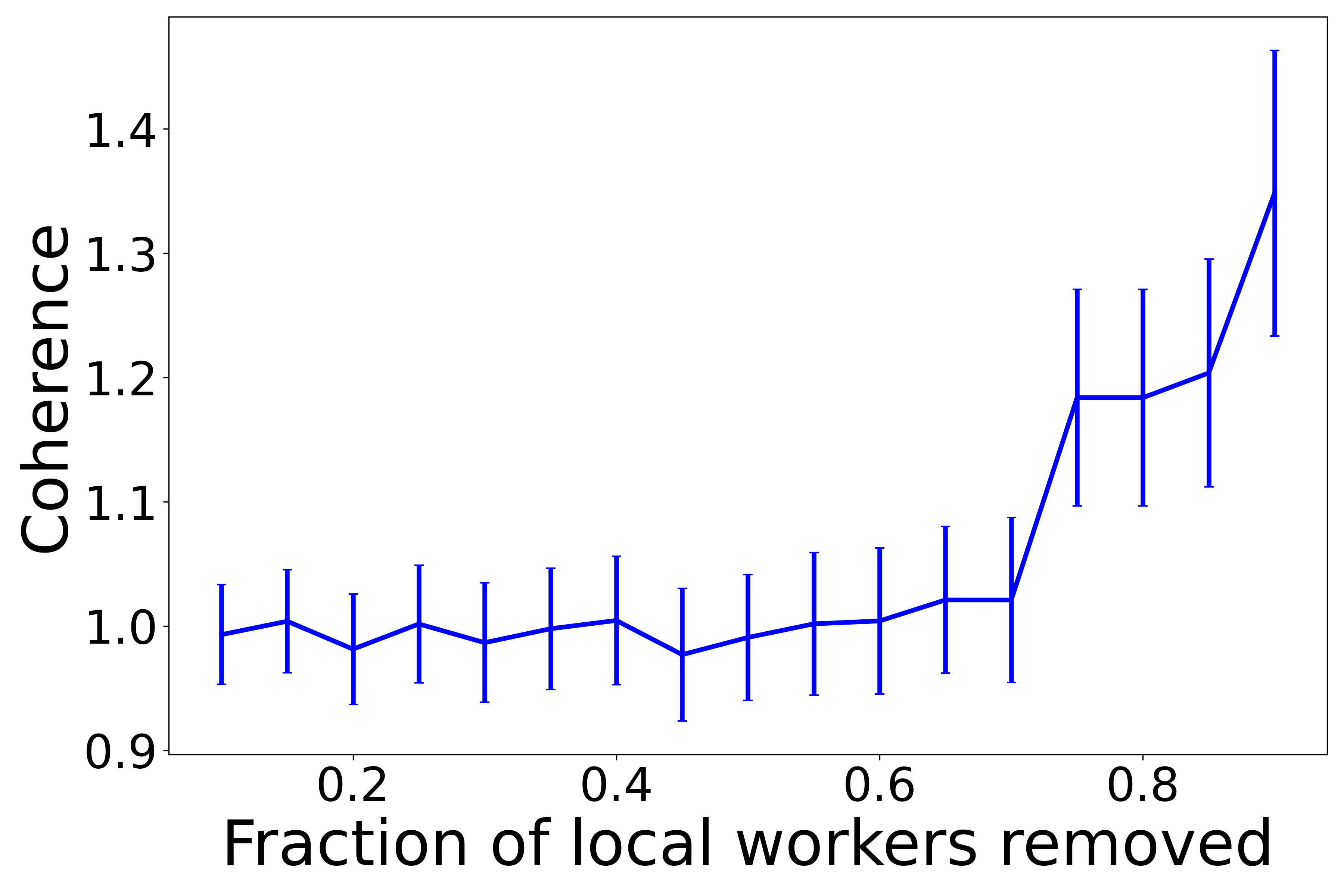} 
    \includegraphics[width=0.4\linewidth]{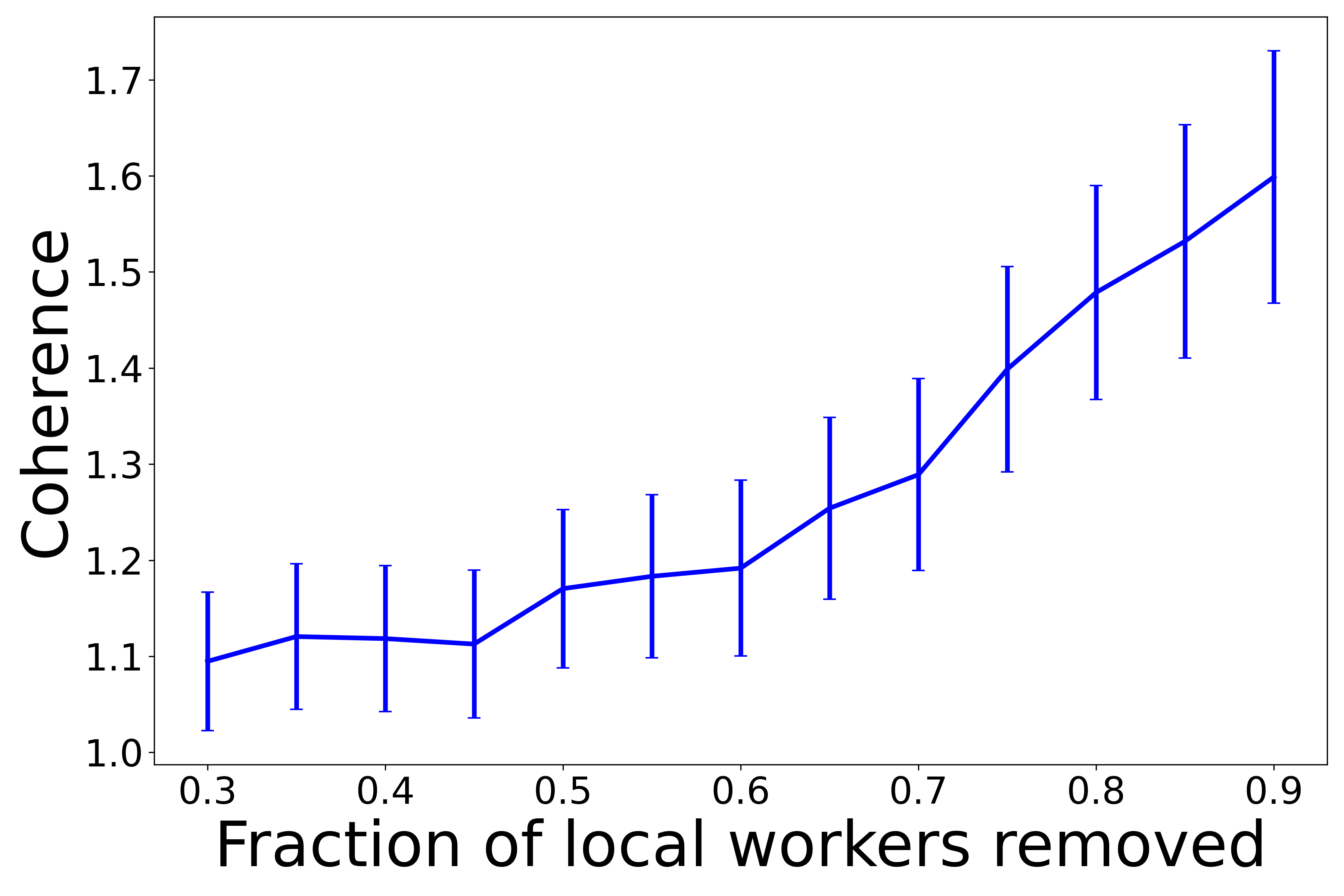} 
    \caption{\textbf{Tradable versus nontradable occupations.} Estimated average coherence across cities in the US for progressively less nontradable occupations. Left panel: census data, right panel BLS data. The data has been aggregated taking the time average over all available years.
    In both panels coherence rises as  occupations become less nontradable.
    }
    \label{fig:tradables_si}
\end{figure}

Fig.~\ref{fig:tradables_si} shows how average coherence in the US urban system changes as we restrict the sample of occupations used in the calculations to  increasingly less nontrable (i.e., more tradable) occupations. Moving from left to right on the horizontal axis, we progressively drop more occupations, by increasing the maximum acceptable $NT$ score. The figures plot coherence averaged across the  time windows considered in the main text  against the share of dropped workers for  occupations in census data (left panel) and in BLS data (right panel).

The census data manifest a sharp transition in coherence after removing the $70\%$ workers in the most nontradable occupations. We therefore use this threshold to distinguish between tradable and nontradable occupations. In the BLS data, no such sharp transition point exists, but the inflection stretches over an interval around the same 70\% value. Therefore, we use  the same cut-off in the BLS data.

\section{Alternative relatedness metrics}
\label{sec:alternative}
In this section we test the robustness of the paper's main results to changes in the definition of occupational relatedness. We explore three different types of relatedness metrics that use different types of data as an input.

\paragraph{Labor flows.}
We create labor flows between occupations using linkages between records of individuals across different census waves developed by Protzer and colleagues\cite{protzer2024new}. This allows us to follow a large number of individuals across census waves and to observe how many individuals change occupations between two waves. To do so, for each pair of sequential decades\footnote{Note that due to the missing 1890 census, 1880 labor flows are taken between 1880 and 1900, crossing two full decades instead of one.}, we count the number of individuals who were employed in occupation $o$ in one decade and subsequently in occupation $o'$ in the next. We collect these labor flows in  matrix $F$, with elements $F_{o o'}$. The resulting metric is used to derive the census results in the main text.

\paragraph{Industrial co-occurrences.}
Industrial co-occurrences measure the extent to which different occupations appear in the same industries. 
To be precise, we count the number of industries in which both occupations of a pair of occupations $(o,o')$ are present to obtain a co-occurrence matrix, $K^{ind}_{o,o'}$. We use this matrix to calculate a relatedness matrix, by replacing $F_{oo'}$ in eq.~(\ref{eq:est_prox}) by $K^{ind}_{oo'}$.    
The resulting occupational relatedness metric is used in the analysis of the  BLS data in the main text.

\paragraph{Geographical co-occurrences (co-agglomeration).}
Following the original approach introduced by Hidalgo and colleagues\cite{hidalgoProductSpaceConditions2007}, we also explore geographical co-agglomeration patterns to measure relatedness. That is, we count how often two occupations are present in the same cities and collect such co-occurrences in matrix $K^{geo}_{oo'}$. Co-occurrences are then converted into a relatedness matrix by replacing $F_{oo' }$ with $K^{geo}_{oo'}$ in eq.~(\ref{eq:est_prox}). 
\vspace{\baselineskip}

\noindent These different measures are likely to emphasize different types of relatedness. Labor flows most likely reveal similarities in skill requirements\cite{neffkeSkillRelatednessFirm2013}, whereas co-occurrence  measures will capture different types of economies of scope. For instance, co-occurrences in industries reflect benefits of combining different types of skills in production processes, whereas co-occurrences at the city level will in addition shed light on agglomeration benefits, such as those derived from labor market pooling effects.

\begin{figure}
    \centering
    \includegraphics[width=0.8\linewidth]{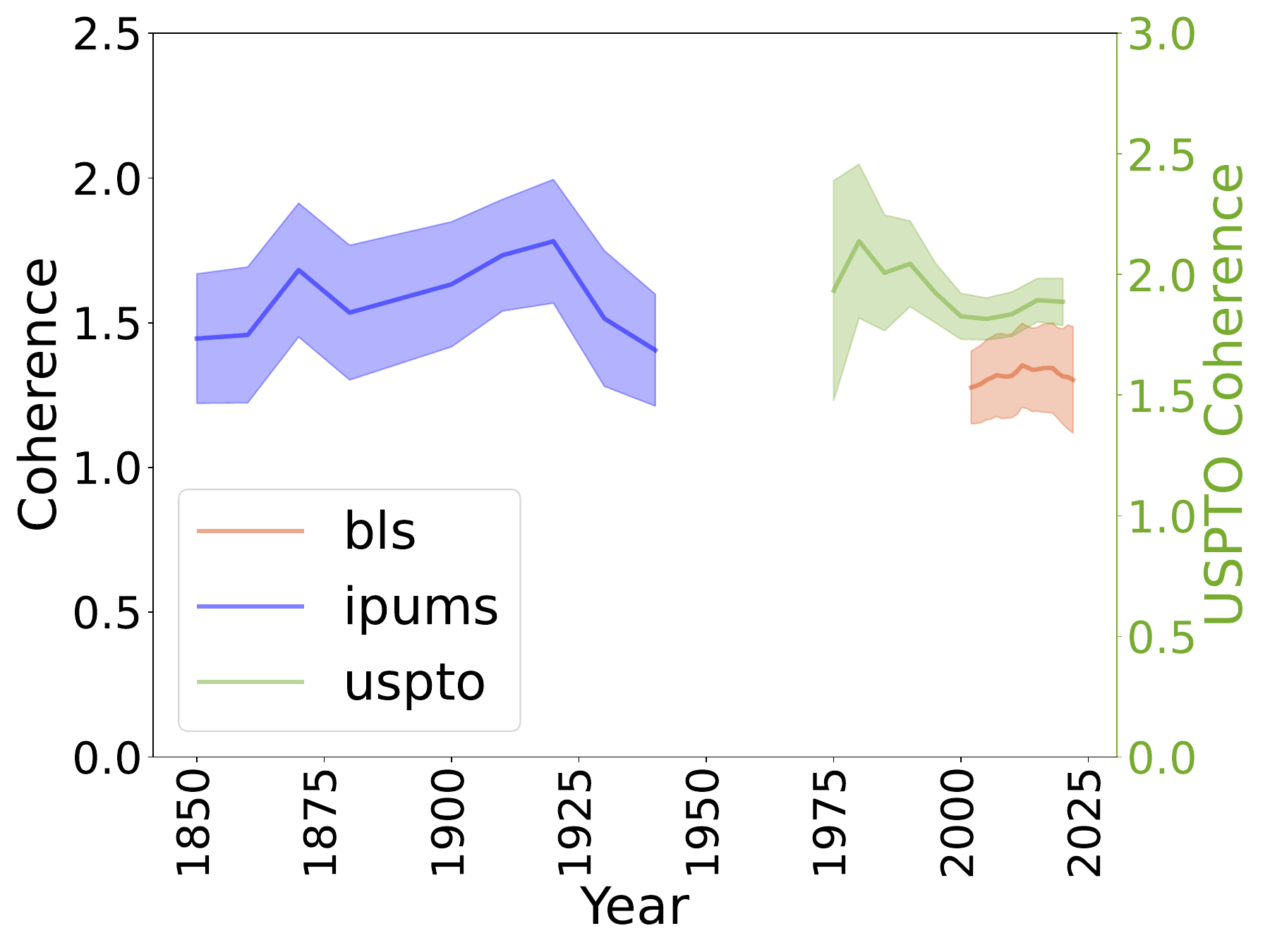} 
    \caption{\textbf{Coherence.}  Mean coherence across US cities with 95\% confidence intervals. Coherence is calculated as in eq.~(\ref{eq:coherence_1}) using employment data from the US census (1850-1940, blue line) and BLS (2002-2022, orange line), and patent data from the USPTO (1980-2020, green line). Values are normalized by dividing by system-level coherence. The proximity matrices are based on city-level co-occurrences.}
    \label{fig:coherence_coocc}
\end{figure}

Fig.~\ref{fig:coherence_coocc} replicates the results in the main text with the alternative relatedness metrics based on co-agglomeration patterns. The upper panel shows that coherence remains more or less unchanged across time periods. The lower panel shows that the elasticity of coherence with respect to city size is constant across time periods and datasets. Moreover, a Wald test that tests whether the estimated elasticities are the same in each year and data set and equal to the average elasticity observed in Fig.~\ref{fig:slopes} of the main text cannot be rejected at any conventional level (p-value: 0.6).     

\section{Variety, Coherence and Capabilities}\label{sec:Hillnumbers}



Diversity is a key concept in research in ecology that describes ecosystems in terms of their composition of species~\cite{macarthurPatternsSpeciesDiversity1965}. 
Methodological work in this area~\cite{vandamDiversityItsDecomposition2019,stirlingGeneralFrameworkAnalysing2007,raoDiversityDissimilarityCoefficients1982}  distinguishes among three fundamental aspects of diversity:

\textbf{Variety}: This describes to the number of distinct species in the ecosystem. A typical metric would be a simple count of the different species in the ecosystem.

\textbf{Balance}: This refers to the distribution or allocation of specimens across the ecosystem's species. Metrics include the Herfindahl-Hirschman Index and entropy.

\textbf{Disparity}: This involves assessing the dissimilarity among species. It examines the question of ``how distinct or dissimilar are different species from each other?''.

Each of these properties -- variety, balance, and disparity -- addresses a specific facet of diversity and different metrics typically focus on one or a combination of these facets. For instance, entropy-based related variety measures~\cite{frenkenRelatedVarietyUnrelated2007} incorporate variety and balance, whereas the density metrics that are often used in economic complexity research~\cite{hidalgoProductSpaceConditions2007,liEvaluatingPrincipleRelatedness2023} incorporate variety and disparity.

In comparison, coherence as defined in eq.~(\ref{eq:coherence}) is similar in spirit to the Rao-Stirling Entropy~\cite{raoDiversityDissimilarityCoefficients1982}, which incorporates all three aspects of diversity: variety, balance and disparity. In fact, slight modifications allow us to turn on or off any of these aspects in the coherence metric to let it mimic specific existing measures. For instance,  imposing $P_{oo'} = \mathbb{I}$ such that the proximity is the identity matrix eliminates disparity. If we furthermore redefine the employment weights in eq.~(\ref{eq:coherence_estimates}) as $E_{co}=\sqrt{E_{co}log(E_{co})}$ we obtain an entropy-based related variety measure\cite{frenkenRelatedVarietyUnrelated2007}. If we instead let  $E_{co}=\frac{M_{co}}{\sum_o{M_{co}}}$, where $M_{co}=1\left(PMI(E_{co}) > 0\right)$, we obtain a density-based variety metric\cite{hidalgoProductSpaceConditions2007}.

\subsection{Elasticity of diversity measures with respect to population}\label{sec:variety_scaling}

\begin{figure}
    \centering
    \includegraphics[width=0.3\linewidth]{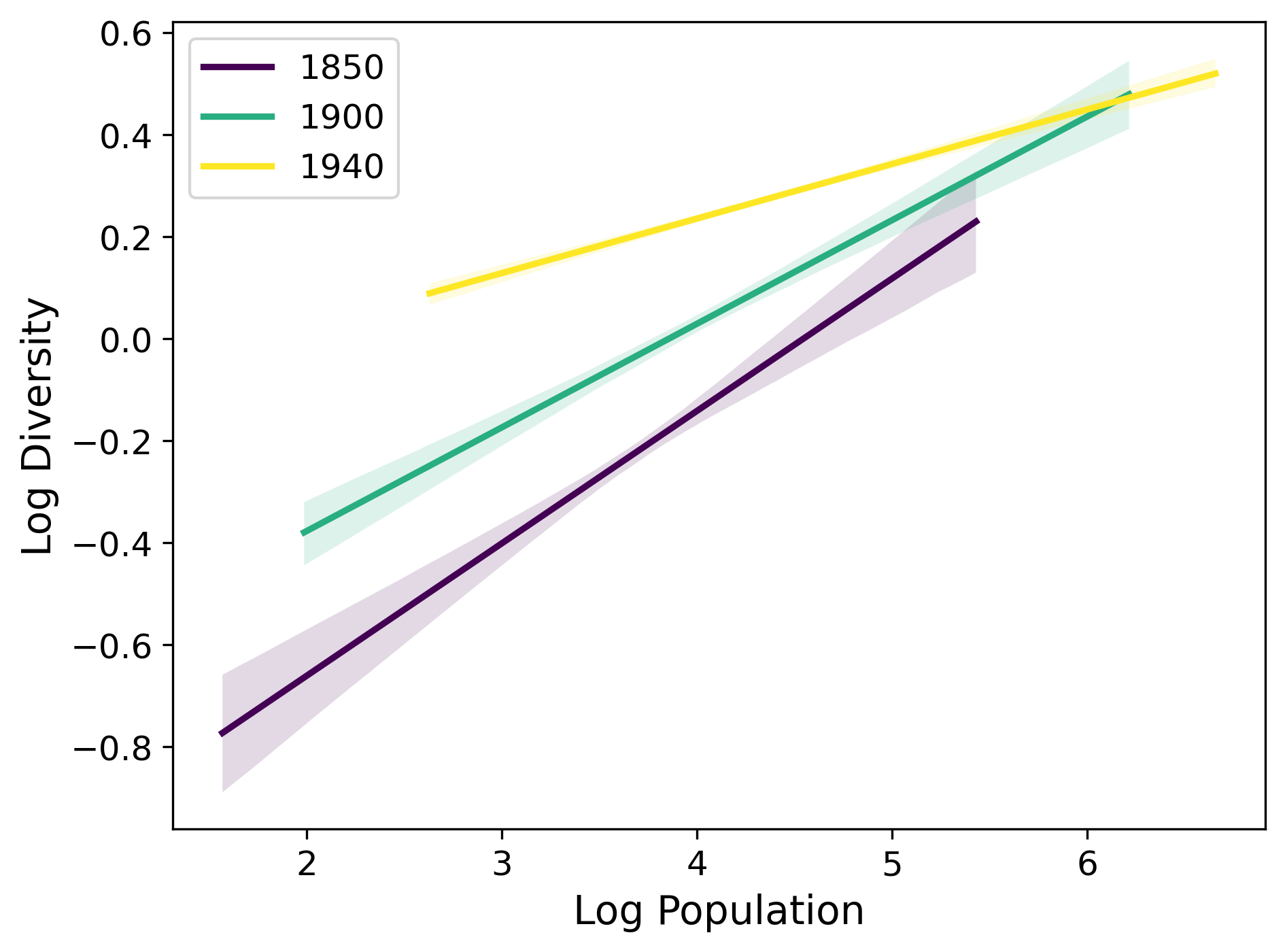}
    \includegraphics[width=0.3\linewidth]{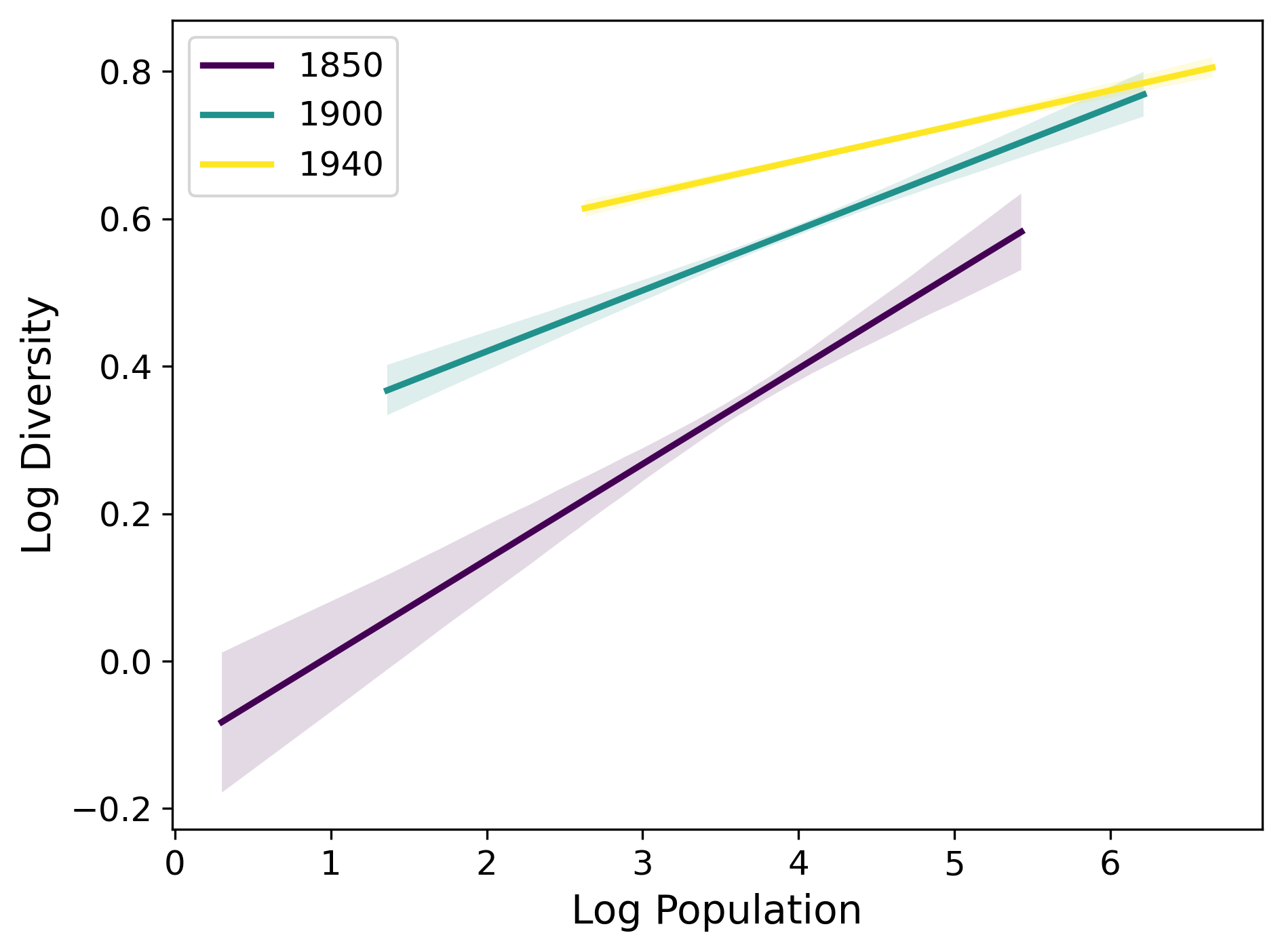}
    \includegraphics[width=0.3\linewidth]{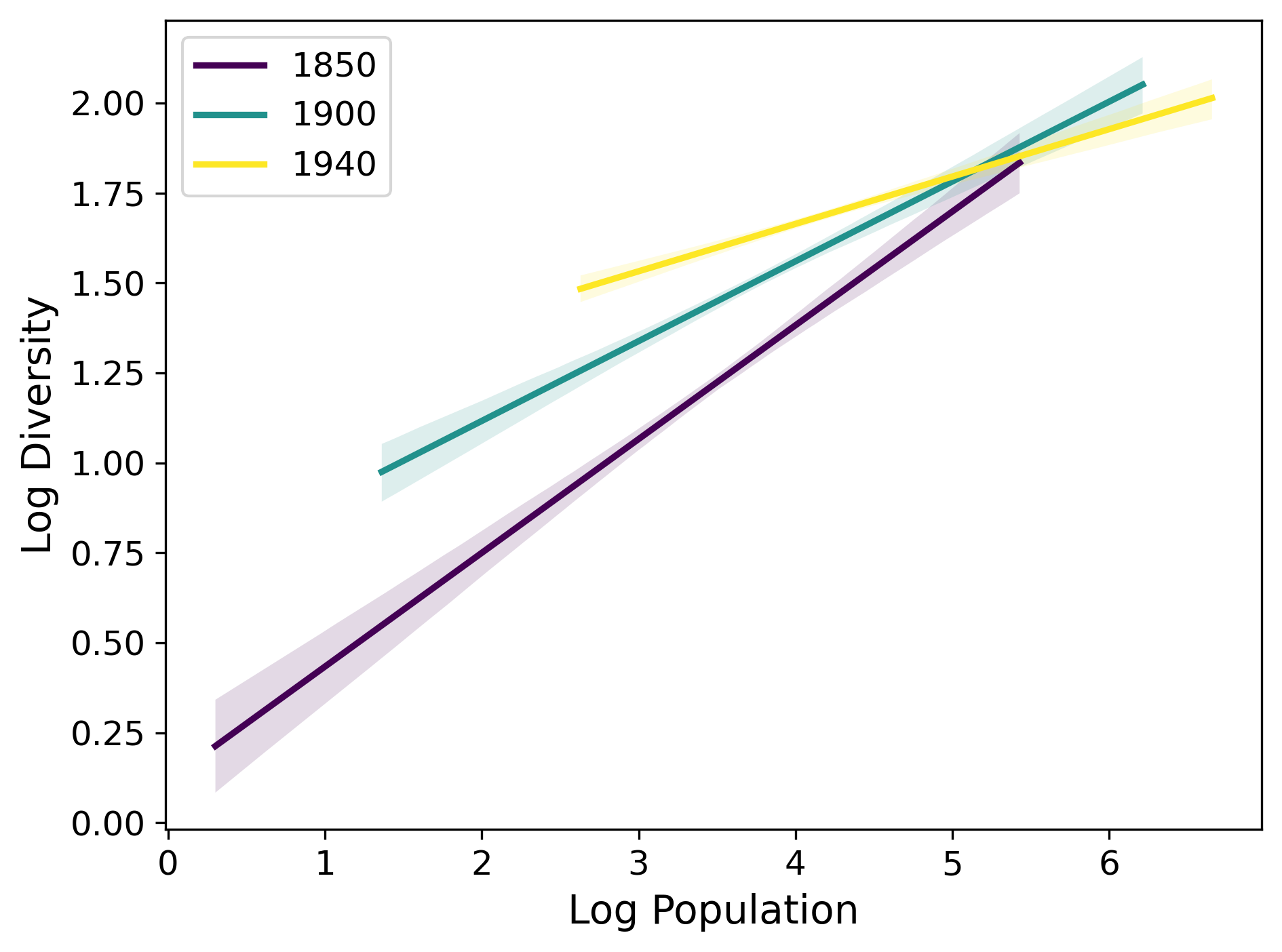} 
    \caption{\textbf{Diversity versus Size.} Left panel: Related variety. Middle panel: entropy, Right panel: RCA-variety, i.e, number of activities with $RCA>1$. Colors refer to different years selected.}
    \label{fig:logvar}
\end{figure}

In the main text, we show that the relation between coherence and city size is unchanging across time periods and data sets. Here, we show that this is not the case for existing measures of diversity. In particular,  Fig.~\ref{fig:logvar} plots the logarithm of diversity against the logarithm of total employment in a city, using as measures of diversity Frenken et al.'s\cite{frenkenRelatedVarietyUnrelated2007} related variety, the entropy of the activity distribution across classes  and a simple count of activities (``RCA-variety'') with $PMI>0$.\footnote{Note that this condition is equivalent to requiring that the Revealed Comparative Advantage (RCA)\cite{balassa1965trade} in the activity exceeds 1.} The plots show that the relation between different diversity measures and population size does not stay constant, but becomes flatter over time. 

\section{Micro simulation}
\label{sec:sim}


To explore mechanisms that could generate the observed relation between coherence and city size, we develop a computational generative model that simulates how productive units select their capabilities. We assume that productive units have two options: either they copy existing capabilities, or they innovate and create a new capability. The model then simulates how the capability base of an urban economy develops with the following key parameters:

\begin{itemize}
    \item \textbf{Capability assignment:} Each new productive unit picks an economic capability through one of two mechanisms:
    \begin{itemize}
        \item With probability $Prob$, the unit adopts an existing capability from to those that are already present in the city.
        \item With probability $1 - Prob$, the individual introduces a new, original capability.
    \end{itemize}
    \item \textbf{Selecting from among the pre-existing capabilities:} As in preferential attachment models, when imitating an existing capability, specific capabilities are selected with a probability that is proportional to their current prevalence in the city.
\end{itemize}

\begin{figure}
    \centering
    \includegraphics[width=0.8\linewidth]{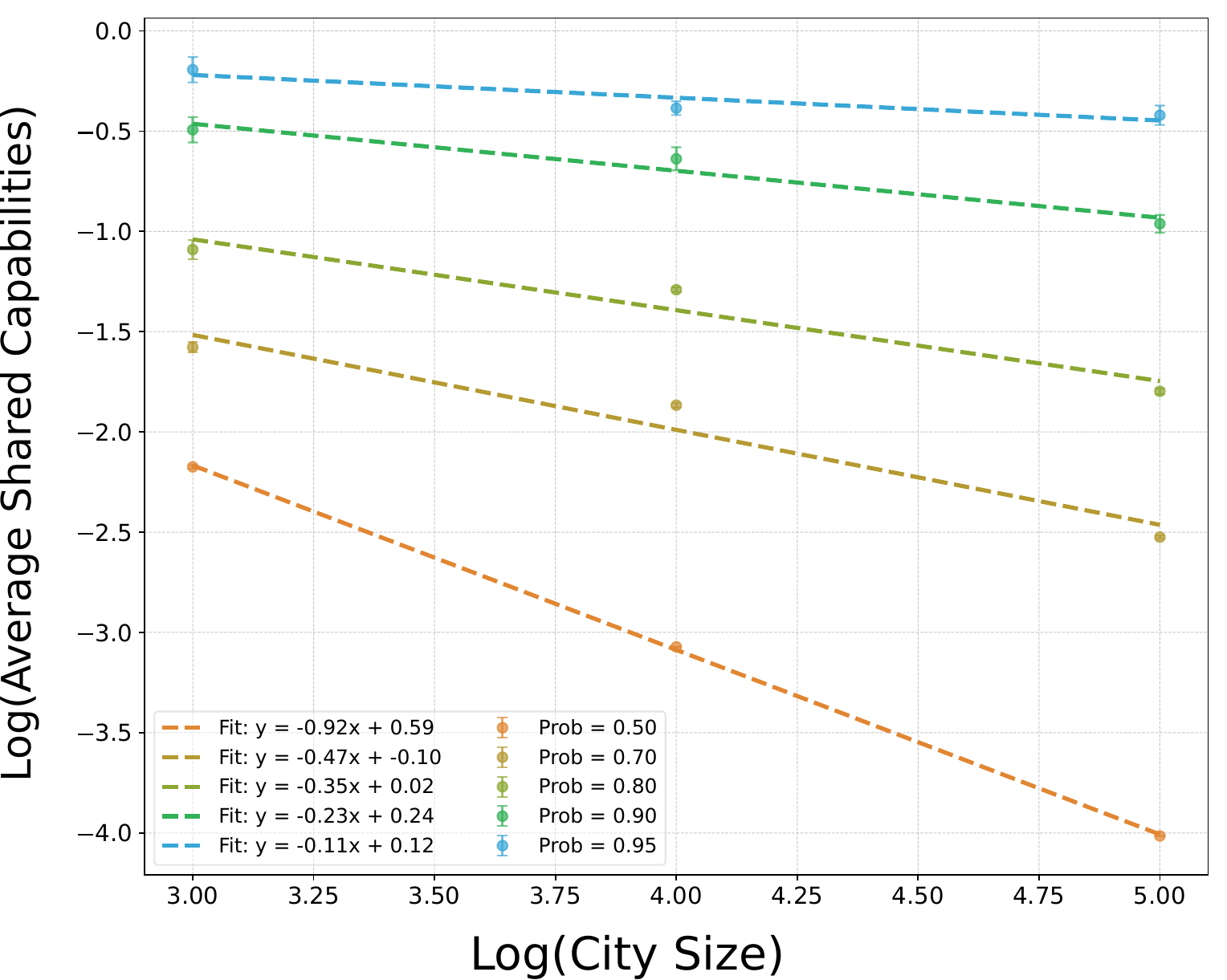} 
    \includegraphics[width=0.8\linewidth]{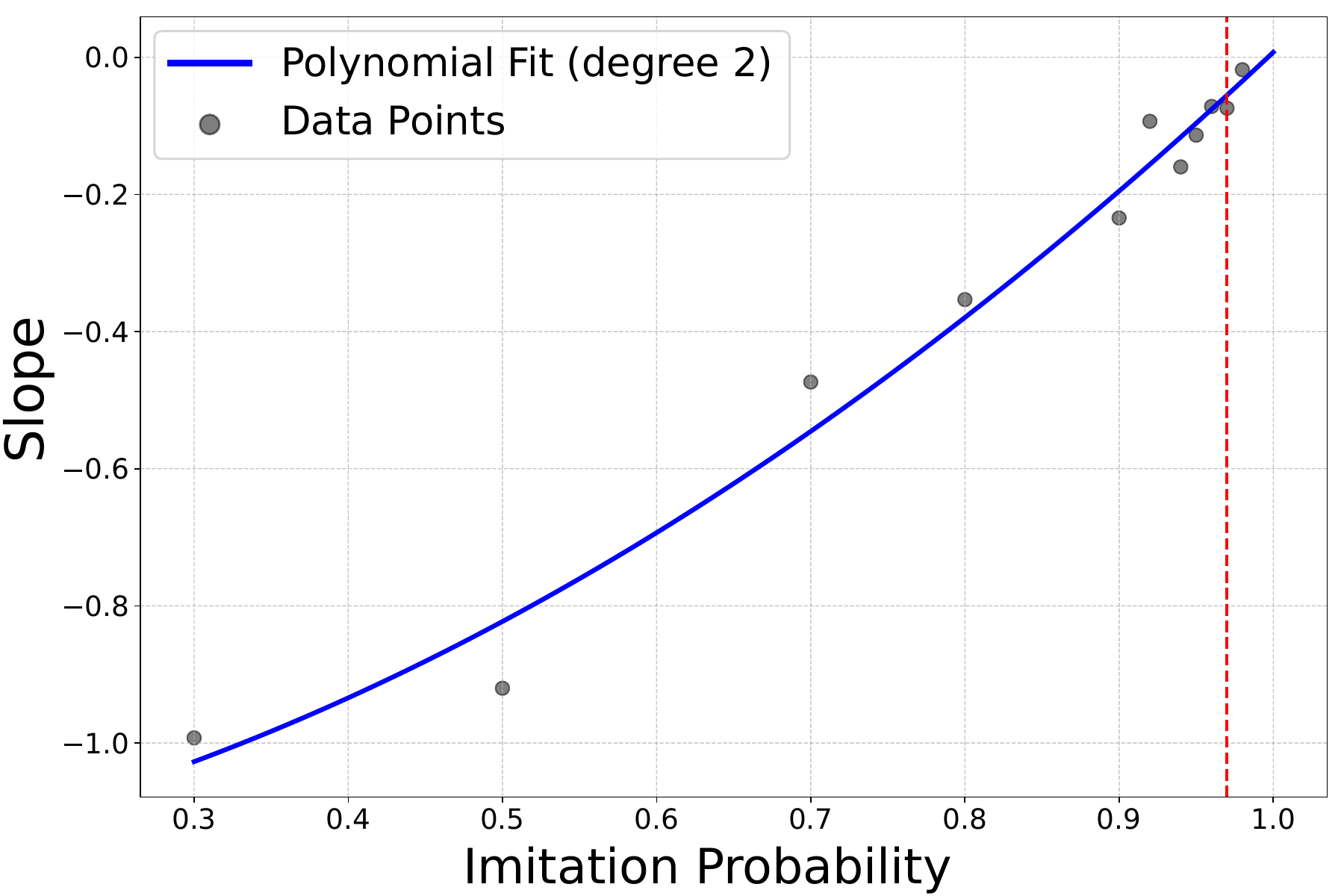} 
    \caption{\textbf{Capability  simulation.} 
The top panel shows how the slope of the size-coherence relationship changes as the imitation probability ($Prob$) increases. The bottom panel plots the estimated elasticities against the probability that units choose to imitate existing capabilities in the city. It shows that an imitation probability of 0.97 (red dotted line) produces a size-coherence slope statistically consistent with our empirical observations.}

    \label{fig:sim}
\end{figure}

As a measure of coherence in this simulation, we estimate the likelihood that two randomly chosen productive units have the same capability. Fig.~\ref{fig:sim} now presents the simulation results under varying innovation probabilities. The left panel shows how the slope of the size-coherence relationship evolves as the probability parameter increases.
The right panel plots the resulting elasticity against the innovation probability. It highlights that an innovation  probability of 0.03 (red dotted line) reproduces the size-coherence slope matching our empirical observations.

This analysis shows that a relatively simple generative model can offer a mechanistic explanation for the self-organization of urban economic structures. It demonstrates how local imitation dynamics can give rise to macroscopic patterns of coherence.

